\documentclass[11pt,a4paper]{article}
\pdfoutput=1

\usepackage[T1]{fontenc}
\usepackage[utf8]{inputenc}
\usepackage{lmodern}

\usepackage{jheppub}
\usepackage{graphicx,amssymb,amsmath,amsfonts}
\usepackage{amsthm}
\usepackage[greek, english]{babel}
\usepackage{hyperref}
\usepackage{cleveref}
\usepackage{subcaption}

\usepackage{braket}
\usepackage{xcolor}

\newcommand{\mn}{{\mu\nu}}
\newcommand{\rs}{{\rho\sigma}}
\newcommand{\ab}{{\alpha\beta}}
\newcommand{\mnrs}{{\mu\nu\rho\sigma}}

\newcommand{\V}{\mathbf}

\newcommand{\p}{{\partial}}

\newcommand{\zb}{{\bar{z}}}
\newcommand{\wb}{{\bar{w}}}
\newcommand{\gzz}{{\gamma_{z\bar{z}}}}

\newcommand{\gww}{{\gamma_{w\bar{w}}}}

\newcommand{\mA}{\mathcal{A}}

\newcommand{\mH}{\mathcal{H}}
\newcommand{\mI}{\mathcal{I}}

\newcommand{\mL}{\mathcal{L}}

\newcommand{\mO}{{O}}

\newcommand{\w}{{\omega}}

\newcommand{\hc}{\text{h.c.}}
\newcommand{\tin}{{\text{in}}}

\newcommand{\tup}{{\text{up}}}

\newcommand{\mmRin}{{\,{}_{-2}R_{l\w}^\text{in}}}
\newcommand{\ppRup}{{\,{}_{+2}R_{l\w}^\text{up}}}

\newcommand{\ppY}{{\,{}_{+2}Y_{lm}}}

\newcommand{\mmY}{{\,{}_{-2}Y_{lm}}}

\newcommand{\mbh}{{M_\text{bh}}}

\makeatother

\title{Supertranslation Hair of Schwarzschild Black Hole: A Wilson Line Perspective}
\author[a]{Sangmin Choi,}
\author[b]{S. Sandeep Pradhan}
\author[a]{and Ratindranath Akhoury}

\affiliation[a]{Leinweber Center for Theoretical Physics, \\
Randall Laboratory of Physics, Department of Physics,\\
University of Michigan, Ann Arbor, MI 48109, USA}

\affiliation[b]{Department of Electrical Engineering and Computer Science, \\
University of Michigan, Ann Arbor, MI 48109, USA}

\emailAdd{sangminc@umich.edu}
\emailAdd{pradhanv@umich.edu}
\emailAdd{akhoury@umich.edu}

\abstract{
	We demonstrate within the quantum field theoretical framework
		that an asymptotic particle falling into the black hole implants soft graviton hair on the horizon,
		conforming with the classical proposal of Hawking, Perry and Strominger.
	A key ingredient to this result is the construction of gravitational Wilson line dressings
		of an infalling scalar field, carrying a definite horizon supertranslation charge.
	It is shown that a typical Schwarzschild state is degenerate, and can be labeled by different soft supertranslation hairs parametrized
		for radial trajectories by the
		mass and energy of the infalling particle and its asymptotic point of contact with the horizon.
	The supertranslation zero modes are also obtained in terms of zero-frequency graviton operators,
		and are shown to be the expected canonical partners of the linearized horizon charge
		that enlarge the horizon Hilbert space.
}

\begin{document}

{
	\maketitle
}

\section{Introduction}\label{sec:introduction}

In the seminal work of Bondi, van der Burg, Metnzer and Sachs \cite{Bondi:1962px,Sachs:1962wk}, it was shown that
	a set of diffeomorphisms that do not vanish at infinity,
	namely the BMS transformations, can act non-trivially on the boundary of asymptotically flat spacetimes.
More recently, Strominger and collaborators \cite{Strominger:2013jfa,He:2014laa} have identified the diagonal subgroup
	of such diffeomorphisms at past and future infinities as an infinite-dimensional
	symmetry group of the S-matrix of perturbative quantum gravity,
	and showed that its associated Ward identities reproduce Weinberg's soft graviton theorem \cite{Weinberg:1965nx}.
This idea was soon applied to QED \cite{He:2014cra,Kapec:2014opa},
	and has led to an extensive analysis of asymptotic symmetries of gauge theories and soft theorems
	\cite{
	Strominger:2013lka,
	Kapec:2015ena, Cachazo:2014fwa, Lysov:2014csa, Strominger:2014pwa, Kapec:2014zla,
	Kapec:2015vwa, He:2015zea, Strominger:2015bla, Dumitrescu:2015fej, Bianchi:2014gla,
	Campiglia:2014yka, Campiglia:2015kxa, Campiglia:2015qka,
	Campiglia:2015yka, Campiglia:2016jdj,
	Laddha:2017vfh,
	Campiglia:2017mua, Campiglia:2017dpg, Campiglia:2017xkp,
	Choi:2017bna,Choi:2017ylo,
	Ashtekar:2018lor,
	Campiglia:2018see,
	Hirai:2018ijc,
	Campiglia:2019wxe,
	He:2019jjk, He:2019pll, Choi:2019rlz}.

One main consequence of this development is that the usual Fock vacuum in a gauge or gravity theory is degenerate,
	and any non-trivial scattering process induces a transition between the degenerate vacua.
What followed was the observation that asymptotic symmetries are related to infrared divergences.
In the presence of massless particles, the quantum S-matrix is ill-defined since all of its matrix elements vanish due to infrared divergences.
It has been shown in \cite{Kapec:2017tkm} that infrared divergence is a consequence of not properly accounting for the vacuum degeneracy.
Indeed, it has been hinted in \cite{Gabai:2016kuf,Choi:2017bna} and shown in \cite{Kapec:2017tkm,Choi:2017ylo}
	that by diagonalizing the conserved charges of the asymptotic symmetry,
	one may re-derive the coherent dressed states of Faddeev and Kulish \cite{Kulish:1970ut,Ware:2013zja},
	which cure the S-matrix elements of infrared divergences \cite{Chung:1965zza,Ware:2013zja,Choi:2017bna}.
Once we diagonalize the conserved charge, the dressing operators can be used to translate one vacuum to another, where each vacuum is
	now labeled with its soft charges.

Building on the fact that the vacuum of asymptotically flat spacetimes is degenerate,
	Hawking, Perry and Strominger \cite{Hawking:2016msc,Hawking:2016sgy} proposed that
	a Schwarzschild black hole can carry an infinite number of soft hair, generated by BMS transformations that act non-trivially on the horizon.
A direct consequence of this is that there exists a large degeneracy of Schwarzschild black holes that have the same mass
	but different soft charge on the horizon.
Drawing from the experience at null infinities, one deduces that a dressed particle that falls into the black hole will induce a
	BMS transformation on the horizon, thereby forming a soft hair.
The conserved BMS charges then has the possibility to constrain the Hawking radiation in a non-trivial manner,
	and thus it was proposed by the authors of \cite{Hawking:2016msc,Hawking:2016sgy} that such soft hairs could have some bearing on
	Hawking's black hole evaporation and information paradox \cite{Hawking:1974sw,Hawking:1976ra}.
Along with some of the earlier work (for example \cite{Flanagan:2015pxa}),
	this led to numerous investigations on the effects of non-trivial diffeomorphisms acting on black hole horizons
\cite{
Averin:2016ybl,Compere:2016jwb,Sheikh-Jabbari:2016lzm,Baxter:2016nml,Compere:2016gwf,
Mao:2016pwq,Averin:2016hhm,Cardoso:2016ryw,Grumiller:2016kcp,Donnay:2016ejv,Tamburini:2017dig,Ammon:2017vwt,
Zhang:2017geq,Mishra:2017zan,
Gomez:2017ioy,
Grumiller:2017otl,Chu:2018tzu,Kirklin:2018wvq,Cvetkovic:2018dmq,Grumiller:2018scv,Chandrasekaran:2018aop,
Averin:2018owq,Donnay:2018ckb,
Compere:2016hzt,Donnay:2015abr}.

Since Faddeev-Kulish (FK) dressings can be used to translate between degenerate vacua of Minkowski spacetime,
	it is reasonable to expect that such dressings on the black hole horizon will implement the soft hair of Hawking, Perry and Strominger.
To construct the Faddeev-Kulish dressings on the horizon,
	a useful ingredient is Mandelstam's gauge-invariant formulation of QED and gravity \cite{Mandelstam:1962mi, Mandelstam:1962us}.
In this formulation, each matter field is dressed with a Wilson line, such that the dressed field as a whole is gauge-invariant.
It was shown by Jakob and Stefanis \cite{Jakob:1990zi} for QED that once this Wilson line is taken to be along the geodesic of
	a free particle at the asymptotic infinity, the Wilson line reproduces Faddeev-Kulish dressings.
This idea was recently adopted in \cite{Choi:2018oel} for QED in a Rindler wedge, to show that a Wilson line along
	the geodesic of a massive particle near the future Rindler horizon implements soft photon hair at the point where the geodesic meets the horizon.
This was interpreted as soft photon hair of Schwarzschild black hole in the near-horizon limit.

In this paper, we use these ideas to demonstrate that asymptotic particles falling into the black hole leave behind a soft graviton hair on the horizon,
	by constructing gravitational dressings on the black hole horizon in the context of perturbative quantum gravity
	in a Schwarzschild background.
This extends the result of \cite{Choi:2018oel} to gravity, with the crucial difference that we work directly in a Schwarzschild background
	instead of a Rindler wedge.
To this end, we quantize the metric perturbation as in \cite{Candelas:1981zv,Jensen:1995qv}, which we review in section \ref{app:quant}.
We observe that the work of Jakob and Stefanis \cite{Jakob:1990zi} extends to gravity in flat background without difficulty,
	and adopt Mandelstam's point of view \cite{Mandelstam:1962us} to construct Faddeev-Kulish dressings as gravitational Wilson lines
	along the geodesic of a massive, radially infalling scalar matter field.
The gravitational Wilson line in curved background is taken to be a straightforward generalization of that in flat background,
	see \cite{Ware:2013zja} for instance.
It is shown that the dressing thus constructed carries a definite soft supertranslation charge parametrized by the mass and energy
	of the matter particle being dressed, in accordance with the case of flat spacetime.
More explicitly, it will be seen that the dressing operator for a radially infalling particle
	of mass $m$ and energy $E$ acting on a black hole state with no hair, creates a black hole with soft hair.
This state is characterized by the energy $E$ and mass $m$ through the ratio $m^2/E$ and by the spherical angles on the future boundary of the future horizon.
This should be contrasted with the case of soft hair on the asymptotic infinities $\mI^\pm$,
	where the corresponding state is labeled by a charge parametrized only by the momentum of the asymptotic particle (see section \ref{sec:flat_charge}).

Our result provides an explicit construction of the soft graviton hair and horizon zero modes of Hawking, Perry and Strominger
	\cite{Hawking:2016msc,Hawking:2016sgy} from the point of view of quantum field theory.
Another outcome of the Wilson line approach of this paper
	is that it illustrates a connection between the asymptotic symmetries and supertranslation charges to the recent discussion of generalized 
	global symmetries \cite{Gaiotto:2014kfa}.
This is outlined in the discussion section.
Moreover, the structure of zero modes and Wilson line dressings suggests subtle connection with analogs of the scalar memory of
	Satishchandran and Wald \cite{Satishchandran:2019pyc} at null infinities of asymptotically flat spacetimes.
We leave further investigation in this direction for future work.

The paper is organized as follows.
In section \ref{sec:dressing}, we first show how gravitational Faddeev-Kulish dressings in flat spacetime can be understood as Wilson lines
	as in \cite{Jensen:1995qv}.
We will use this method in section \ref{sec:H} to construct the dressings on the Schwarzschild black hole horizon.
In section \ref{sec:st_charge} we demonstrate that the dressings carry a definite soft supertranslation charge.
In this process, we observe that the supertranslation zero modes of \cite{Hawking:2016sgy} are naturally derived in terms of
	zero-energy graviton operators.
In section \ref{sec:implant}, we show that the dressings can be used to implement soft hair on the black hole horizon.
A discussion of our results is given in section \ref{sec:discussion}.
Appendix \ref{app:nnc} comprises our notation and conventions (of the Newman-Penrose formalism, among others).
In appendix \ref{app:quant} we review the quantization of metric perturbation of \cite{Candelas:1981zv,Jensen:1995qv}.
Since we make use of spin-weighted spherical harmonics extensively, a short review of their properties is given in appendix \ref{app:swsh}.
Appendix \ref{app:useful} contains the Christoffel symbols of the Schwarzschild metric in Eddington-Finkelstein coordinates,
	and also derives some useful properties of spin-2 spherical harmonics that will be used in the main text.
In appendix \ref{app:Qminus}, we derive the supertranslation charge on the past Schwarzschild horizon
	using similar methods as in \cite{Hawking:2016sgy}.
Some technical details are delegated to appendices \ref{app:magnetic} and \ref{app:unfold}.

\section{Review of gravitational dressings at infinity in flat spacetime} \label{sec:dressing}

In this section, we review the gravitational FK dressings in flat spacetime \cite{Ware:2013zja}
	and its Wilson line representation \cite{Mandelstam:1962us,Jakob:1990zi}.
We also review how the dressings carry a definite supertranslation charge \cite{Choi:2017bna,Choi:2017ylo}.
These results will be central to our construction of dressings on the Schwarzschild horizon.

\subsection{Dressing as a Wilson line}

Mandelstam  \cite{Mandelstam:1962us} formulated a method for quantizing gravity using path-dependent
	but coordinate-independent variables.
This involves quantizing of the curvature tensor field directly in a path-dependent way instead of the
standard quantization of the metric tensor field (gauge field). 
Consider the interaction between a scalar field and gravitational field.   
Let us consider a small perturbation with respect to the flat spacetime,
\begin{align}
	g_\mn(x) = \eta_\mn + \kappa h_\mn(x),
	\label{flat_metric_decomp}
\end{align}
where $\eta_\mn = \text{diag}(-1,1,1,1)$ is the flat metric and $\kappa^2 = 32\pi G$ with Newton's constant $G$.
The prescription of writing a path-dependent variable $A(x,P)$ in terms of a coordinate-dependent variable $a(x)$
	is given in \cite{Mandelstam:1962us}, which reads
\begin{align}
	A(x,P) = a(x) + \frac{i\kappa}{4}\int^x_P dz^\lambda
		\left\{
			\frac{\p h_{\mu\lambda}(z)}{\p z^\nu}
			- \frac{\p h_{\nu\lambda}(z)}{\p z^\mu}
		\right\}\left[J^\mn(z),a(x)\right]
		- \frac{\kappa}{2}\int^x_P dz^\lambda h_{\mu\lambda}(z)\frac{\p a(x)}{\p x_\mu},
\end{align}
to first order in $\kappa$.
Here $J_\mn(z)$ is the angular momentum operator about $z$ in the $\mn$-plane.
Let us consider the case where the variables are scalar fields of mass $m$, and write $A(x,P)=\Phi(x,P)$ and $a(x) = \phi(x)$.
Since our focus is on supertranslation charges, and the angular
momentum term is sub-leading, we arrive at the following expression for
the leading order:
\begin{align}
	\Phi(x,P)
	&= \phi(x) - \frac{\kappa}{2}\int^x_P dz^\lambda h_{\mu\lambda}(z)\frac{\p \phi(x)}{\p x_\mu}
	\\ &=
		\left\{
			1 - \frac{i \kappa}{2}\int^x_P dz^\lambda h_{\mu\lambda}(z)\left(-i\p^\mu\right)
		\right\}
		\phi(x)
	\\ &= W_1(x,P)\phi(x),
\end{align}
where the operator $W_1(x,P)$ is defined as
\begin{align}
	W_1(x,P) \equiv
		1 - \frac{i \kappa}{2}\int^x_P dz^\lambda h_{\mu\lambda}(z)\left(-i\p^\mu\right).
\end{align}
This can be interpreted as an operator that dresses the scalar field.  
When the scalar field is quantized, we have the standard expansion
\begin{align}
	\phi(x) = \int \frac{d^3p}{(2\pi)^3}\frac{1}{2E_\V{p}}
		\left\{
			a(\V p) e^{ip\cdot x} + a^\dagger(\V p) e^{-ip\cdot x}
		\right\},
\end{align}
where $E_\V{p}^2 = \V p^2 + m^2$ and the creation/annihilation operators $a$, $a^\dagger$ satisfy the commutation relation
\begin{align}
	\left[a(\V p), a^\dagger(\V q)\right] = (2\pi)^3 (2 E_\V{p}) \delta^{(3)}(\V p-\V q).
\end{align}
The expansion of $\phi(x)$ shows that dressing each scalar field with the operator $W_1(x,P)$ is essentially equivalent to
	dressing each operator $a(\V p)$ with $W_1(\V p;x, P)$ defined as
\begin{align}
	W_1(\V p; x, P) = 1-\frac{i \kappa}{2}\int^x_P dz^\lambda h_{\mu\lambda}(z) p^\mu,
\end{align}
where $p^\mu = (E_p, \V p)$, and each operator $a^\dagger(\V p)$ with $W_1^\dagger(\V p; x, P)$.
Notice that this is the first-order approximation of the Wilson line $\mathcal W(\V p; x, P)$, defined as
\begin{align}
	W(\V p; x, P) = \mathcal{P} \exp\left\{-\frac{i \kappa}{2}\int^x_P dz^\lambda h_{\mu\lambda}(z) p^\mu\right\}
\end{align}
with path-ordering $\mathcal{P}$.
Taking the path $P$ to be a trajectory of a free particle with velocity $v=p/m$ allows us to parametrize
	$z = x + v\tau$ and write
\begin{align}
	W(\V p;x)
	= \mathcal{P} \exp\left\{-\frac{i \kappa}{2}\int^0_{-\infty} d\tau\, p^\mu v^\nu  h_\mn(x + v\tau)\right\}.
\end{align}
Notice the similarity between this Wilson line operator defined in gravity with that defined in QED \cite{Jakob:1990zi, Choi:2018oel}.
Next we consider the metric perturbations.

The Einstein's equations $G_\mn = \frac{\kappa^2}{2}T_\mn$ in terms of the perturbation reads \cite{Donoghue:2017pgk}
\begin{align}
	O_{\mn \rs}h^\rs = \frac{\kappa}{2}T_\mn,
\end{align}
where
\begin{align}
	{O^\mn}_\rs =
		\left(
			\frac{1}{2}\delta^\mu_\rho \delta^\nu_\sigma
			+ \frac{1}{2}\delta^\nu_\rho \delta^\mu_\sigma
			- \eta^\mn\eta_\rs
		\right)\square
		- \frac{1}{2}\left(
			\delta^\mu_\rho \p^\nu \p_\sigma
			+ \delta^\mu_\sigma \p^\nu \p_\rho
			+ (\mu\leftrightarrow\nu)
		\right)
		+ \eta_\rs \p^\mu\p^\nu
		+ \eta^\mn \p_\rho\p_\sigma.
\end{align}
The operator $O_{\mn\rs}$ is not invertible, so we must fix the gauge using the harmonic gauge condition
\begin{align}
	\p_\mu {h^\mu}_\nu - \frac{1}{2}\p_\nu h = 0,
\end{align}
after which we obtain the retarded Green's function \cite{Donoghue:2017pgk}
\begin{align}
	G^\text{ret}_{\mn\rs}(x) = \frac{1}{2}I_\mnrs\frac{1}{4\pi |\V x|}\delta(|\V x| - x^0),
\end{align}
where $I_\mnrs = \eta_{\mu\rho}\eta_{\nu\sigma} + \eta_{\mu\sigma}\eta_{\nu\rho} - \eta_\mn\eta_\rs$, satisfying
\begin{align}
	{O_\mn}^\rs G^\text{ret}_{\rs\alpha\beta}(x - y) = \frac{1}{2}\V 1_{\mn\alpha\beta}
		\delta^{(4)}(x-y),
\qquad
	G^\text{ret}_\mnrs (x-y) = 0 \quad\text{if $x^0 < y^0$}.
\end{align}
Here the ``identity" tensor $\V 1_{\mn\alpha\beta}$ is defined as
\begin{align}
	\V 1_{\mn\alpha\beta} = \frac{1}{2}(\eta_{\mu\alpha}\eta_{\nu\beta}+\eta_{\mu\beta}\eta_{\nu\alpha}).
\end{align}
Using the retarded Green's function, we may express metric
perturbations in terms of free fields as 
\begin{align}
	h_{\mu \nu}(x)=h_{\mu \nu}^\tin(x)+2\int d^4y \,G^\text{ret}_\mnrs(x-y) T^\rs(y),	
\end{align}
where $h_\mn^\tin$ is the free field satisfying $O^\mnrs h_\rs = 0$,
and $T^\mn$ is the energy-momentum tensor of the scalar field
which is given by (see eq. (4.2) in \cite{Ware:2013zja})
\begin{align}
  T^{rs}(y)= \kappa \int d^3 \mathbf{p} \rho(\mathbf{p}) \frac{p^r p^s}{2p_0}
  \delta^{(3)}\left( \mathbf{y}-\frac{\mathbf{p}t}{p_0} \right), 
\end{align}
where
$\rho(\mathbf{p})$ is the (unintegrated) number operator 
$a^{\dagger}(\mathbf{p}) a(\mathbf{p})$. 
It follows that the Wilson line $W(\V p; x)$ can be  written in terms
of free field and the energy-momentum tensor as 
\begin{align}
	W(\V p; x) &= \mathcal{P} \exp\left\{
		-\frac{i\kappa}{2}\int^0_{-\infty} d\tau v^\mu p^\nu h^\tin_\mn(x+v\tau)
	\right\}
	\\&\qquad\times\exp\left\{
		-i\kappa\int^0_{-\infty} d\tau v^\mu p^\nu \int d^4y \,G^\text{ret}_\mnrs(x+v\tau-y) T^\rs(y)
	\right\}
\end{align}
Next let us recall the  Wick's ordering theorem, given by
\begin{align}
T \exp \left[ -i \int dt\, H_I(t) \right] = \exp\left[ -i \int H_I(t) \right] 
\exp \left[-\frac{i}{2} \int dt\int ds\, \theta(t-s) [H_I(t),H_I(s)] \right],
\end{align}
where $\theta$ is the step function.
Using this, we can express the Wilson line operator as \cite{Jakob:1990zi}
\begin{align} 
W(\V p;x) &= \exp \left\{ - \frac{i \kappa}{2} \int_{-\infty}^0 d \tau\, v^{\mu}
	p^{\nu} h_{\mu \nu}^{\tin}(x+v\tau) \right\} \times \text{(phases)}
	\\ &= \widetilde W(\V p;x) \times \text{(phases)}.
\end{align}
We will not concern ourselves with the phases, and focus on the first factor,
\begin{align}
	\widetilde W(\V p;x)\equiv \exp \left\{ - \frac{i \kappa}{2} \int_{-\infty}^0 d \tau\, v^{\mu}
		p^{\nu} h_{\mu \nu}^{\tin}(x+v\tau) \right\}.
\end{align}
Using the standard mode expansion of the asymptotic in-field,
\begin{align}
	h^\tin_\mn(x) = \int \frac{d^3k}{(2\pi)^3}\frac{1}{2\w}
		\left[
			a_\mn(\V k) e^{ik\cdot x}
			+ a_\mn^\dagger(\V k) e^{-ik\cdot x}
		\right],
\end{align}
where $\w \equiv \left|\V k\right|$, we obtain
\begin{align}
	\int_{-\infty}^0 d \tau\, v^{\mu}p^{\nu} h_{\mu \nu}^{\tin}(x+v\tau)
	&= \int_{-\infty}^0 d \tau\, v^{\mu}p^{\nu}
		\int \frac{d^3k}{(2\pi)^3}\frac{1}{2\w}
		\left[
			a_\mn(\V k) e^{ik\cdot (x+v\tau)}
			+ a_\mn^\dagger(\V k) e^{-ik\cdot (x+v\tau)}
		\right]
	\\ &= i\int \frac{d^3k}{(2\pi)^3}\frac{1}{2\w}
		\frac{p^\mu p^\nu}{p\cdot k}
		\left[
			a_\mn^\dagger(\V k)e^{-ik\cdot x} - a_\mn(\V k) e^{ik\cdot x}
		\right],
\end{align}
where we used $p=mv$ and the boundary condition
\begin{align}
	\int_{-\infty}^0 d\tau\,e^{ik\cdot v\tau} = \frac{1}{ik\cdot v}.
\end{align}
Recall that $x$ is the position of the scalar field which is dressed by the Wilson line.
The dressing of a scalar field at the past time-like infinity can be obtained by taking the limit $x^0 \to -\infty$.
Due to the factors $e^{\pm ik\cdot x}$, by the Riemann-Lebesgue lemma only the leading soft particles contribute to the integral.
Following \cite{Kulish:1970ut}, we implement this using an infrared function $\phi(\w)$ that has support in a small neighborhood of
	$\w=0$ and $\phi(0)=1$,
\begin{align}
	W(\V p) &= \lim_{x^0\to-\infty} \widetilde W(\V p;x)
	\\ &= \exp\left\{
		\frac{\kappa}{2}\int \frac{d^3k}{(2\pi)^3}\frac{1}{2\w}
		\frac{p^\mu p^\nu}{p\cdot k}\phi(\w)
		\left[
			a_\mn^\dagger(\V k) - a_\mn(\V k) 
		\right]
		\right\},
	\label{dressing_flat}
\end{align}
which is, up to a unitary transformation, identified with the Faddeev-Kulish dressing of gravity \cite{Ware:2013zja}.
The dressings may be interpreted as Wilson line punctures \cite{Blommaert:2018oue,Blommaert:2018rsf,Choi:2018oel} on the spacetime boundary.

\subsection{Wilson line punctures and boundary charges}\label{sec:flat_charge}

In this subsection, let us review how the dressing \eqref{dressing_flat} carries BMS supertranslation charge \cite{Choi:2017bna,Choi:2017ylo}.
We will work with the asymptotically flat metric, and use the Bondi coordinates $(v,r,z,\bar{z})$ \cite{He:2014laa},
\begin{align*}
	ds^2 &= -dv^2 +2dvdr +2r^2\gamma_{z\zb}dzd\zb
	\nonumber\\&\quad+\frac{2m_B}{r}dv^2+rC_{zz}dz^2+rC_{\zb\zb}d\zb^2-2V_zdvdz-2V_\zb dvd\zb+\cdots,
\end{align*}
where $\gamma_{z\zb}=2/(1+z\zb)^2$ is the 2-sphere metric,
	$m_B$ is the Bondi mass aspect, and $V_z=\frac{1}{2}D^zC_{zz}$ with $D_z$ the covariant derivative on the 2-sphere.
The first line corresponds to the flat metric.

We start by using \eqref{flat_metric_decomp} to write the radiative data $C_{zz}$ as (see for example \cite{He:2014laa})
\begin{align}
C_{zz}(v,z,\bar{z}) &= \kappa \lim_{r \rightarrow \infty}  \frac{1}{r} h_{zz}(r,v,z,\bar{z}) \\
&=\kappa \lim_{r \rightarrow \infty}  \frac{1}{r} \partial_z x^{\mu} \partial_z x^{\nu} h_{\mu \nu } \\
&= -\frac{i \kappa \gzz}{8 \pi^2}  \int_0^{\infty} d\w \left[
a_+(-\w \hat{\mathbf{x}}_z) e^{-i \w v} - a_{-}^{\dagger} (-\w \hat{\mathbf{x}}_z) e^{i\w v} \right].
	\label{Czz}
\end{align}
Taking the limit $v\to-\infty$, we obtain
\begin{align}
	C_{zz}(z,\zb) &\equiv \lim_{v\to-\infty}C_{zz}(v,z,\zb)
		\\ &= \frac{i\kappa\gzz}{8\pi^2}\int_0^{\infty} d\w \left[ a_{-}^{\dagger}(-\w \hat{\mathbf{x}}_z) -a_+(-\w \hat{\mathbf{x}}_z)\right] \phi(\w),
	\label{Czz2}
\end{align}
where we have again used the Riemann-Lebesgue lemma and the infrared function $\phi(\w)$.
Now rewrite the dressing \eqref{dressing_flat} as,
\begin{align}
	W(\V p) &= \exp \left[ \frac{\kappa}{2} \int \frac{d\w d^2z \gamma_{z\bar{z}}}{16 \pi^3} 
\frac{p^{\mu} p^{\nu} }{p \cdot \hat{k}}\phi(\w)
\left(a_{\mu \nu}^{\dagger}(\mathbf{k})
-a_{\mu \nu}(\mathbf{k}) \right) 
\right]  \\
&= \exp \left[ \frac{\kappa}{2} \int \frac{d^2z \gamma_{z\bar{z}}}{16 \pi^3} 
 \left\{ \frac{(p\cdot \epsilon^-)^2 }{p \cdot \hat{k}}
\int d\w\, \phi(\w) \left(a_{-}^{\dagger} (-\w \hat{\mathbf{x}})
-a_{+}(-\w \hat{\mathbf{x}}) \right) \right. \right. \\
&  \hspace{1.5in}  +\left.  \left. \frac{(p\cdot \epsilon^+)^2 }{p \cdot \hat{k}}
\int d\w\, \phi(\w) \left(a_{+}^{\dagger} (-\w \hat{\V x})
-a_{-}(-\w \hat{\V x}) \right) \right\}  \right],
	\label{this_form}
\end{align}
where $\hat k^\mu = (1,\hat{\V k})=(1,-\hat{\V x})$ and we have used
\[
a_{\mu \nu} (\mathbf{k}) =\sum_{r=\pm} \epsilon^{r*}_{\mu \nu}(\mathbf{k}) a_r (\mathbf{k}), 
\]
with graviton polarization tensors $\epsilon^{\pm}_{\mu \nu}(\mathbf{k})=\epsilon^{\pm}_{\mu}(\mathbf{k})
\epsilon^{\pm}_{\nu}(\mathbf{k})$, defined as $\epsilon^{-\mu}(\mathbf{k})=\frac{1}{\sqrt{2}} [z,1,+i,-z]$ and 
$\epsilon^{+\mu} (\mathbf{k})=\frac{1}{\sqrt{2}} [\bar{z},1,-i,-\bar{z}]$.
In this form \eqref{this_form}, we can see that the dressing may be written in terms of $C_{zz}(z,\zb)$ as
\begin{align}
W(\V p) &= \exp \left[ -\frac{i}{4 \pi} \int d^2z \left\{
	\frac{(p\cdot \epsilon^-)^2 }{p \cdot\hat{k}}
	C_{zz}(z,\bar{z}) +
	\frac{(p\cdot \epsilon^+)^2 }{p\cdot \hat{k}}
	C_{\bar{z} \bar{z}}(z,\bar{z}) \right\} \right].
	\label{Wp}
\end{align}
This form will be convenient in computing the dressing's supertranslation charge.

Next we consider the soft BMS supertranslation charge $Q^{\mI^-}_f$, which is given as \cite{He:2014laa}
\begin{align}
	Q^{\mI^-}_f = \frac{4}{\kappa^2}\int dv d^2z \gamma^{z\zb} D^2_zf(z,\zb) \p_v C_{zz}
		= \frac{4}{\kappa^2}\int dv d^2z \gamma^{z\zb} D^2_\zb f(z,\zb) \p_v C_{\zb\zb},
\end{align}
where $f(z,\zb)$ is the 2-sphere function that parametrizes the transformation.
Let us define the operator
\begin{align}
	M_{zz}(z,\zb) &= \frac{4}{\kappa^2}\gamma^{z\zb}\int _{-\infty}^\infty dv\, \p_v C_{zz}(v,z,\zb)
	\\ &= -\frac{1}{ \kappa\pi} \int_0^{\infty} d\w\,\w\delta(\w) \left[
		a_+(-\w \hat{\mathbf{x}}_z) + a_{-}^{\dagger} (-\w \hat{\mathbf{x}}_z)\right],
\end{align}
where we have used the mode expansion \eqref{Czz} and the integral representation
\begin{align}
	\pm 2\pi i\w \delta(\w) = \int_{-\infty}^\infty dv\,\p_v\,e^{\pm i \w v}.
\end{align}
Then, we may write $Q^{\mI^-}_f$ as
\begin{align}
	Q^{\mI^-}_f = \int d^2z\,M_{zz} D^2_zf(z,\zb) 
		= \int d^2z\, M_{\zb\zb} D^2_\zb f(z,\zb).
		\label{Q}
\end{align}
Using \eqref{Czz2}, one can see by direct calculation that the following commutation relation is satisfied:
\begin{align}
	\left[M_{zz}(z,\zb), C_{\wb\wb}(w,\wb)\right]
	&= 
		\frac{i\gww}{4 \pi^3} \int_0^{\infty} d\w \,\w\delta(\w)\frac{(2\pi)^3(2\w)}{\w^2 \gzz}\delta^{(2)}(w-z)
	\\ &= 
		4i \delta^{(2)}(w-z) \int_0^{\infty} d\w \,\delta(\w)
	\\ &= 
		2i \delta^{(2)}(w-z),
\end{align}
where in the last line we used the convention
\begin{align}
	\int_0^\infty d\w\, f(\w)\delta(\w) = \frac{1}{2}f(0).
\end{align}
This commutation relation shows that the operator $\frac{1}{2}M_{zz}$ is the canonical conjugate variable of the radiative mode $C_{zz}$.
It follows from \eqref{Q} that
\begin{align}
	[Q^{\mI^-}_f, C_{zz}] = 2i D_z^2 f(z,\zb),
	\label{comm}
\end{align}
which is the correct commutator between the soft supertranslation charge and the radiative mode $C_{zz}$ \cite{He:2014laa} in the past infinity.

From \eqref{Wp} and \eqref{comm}, we obtain the commutator
\begin{align}
	\left[Q^{\mI}_f, W(\V p)\right] = Q_f(\V p)W(\V p),
	\label{comm2}
\end{align}
where $Q_f(\V p)$ is given in terms of the 2-sphere function $f$ and the momentum $\V p$,
\begin{align}
	Q_f(\V p)
	&=
		\frac{1}{2 \pi} \int d^2z
		\left\{
			\frac{(p\cdot \epsilon^-)^2 }{p \cdot\hat{k}}
				D_z^2f
			+ \frac{(p\cdot \epsilon^+)^2 }{p\cdot \hat{k}}
			D_\zb^2f
		\right\}.
\end{align}
The commutator \eqref{comm2} shows that the FK dressing $W(\V p)$ carries a definite supertranslation charge $Q_f(\V p)$.

\section{Gravitational dressing on the Schwarzschild horizon}\label{sec:H}

\begin{figure}[t]
	\centering
	\includegraphics[height=.25\textheight]{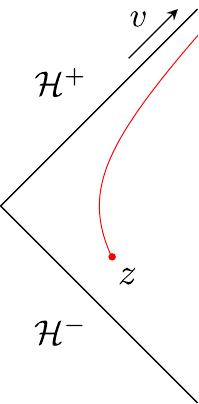}
	\caption{Depiction of a gravitational Wilson line in the Schwarzschild background (marked red in the figure).
		The line extends from the spacetime point $z=(t_0,r_0,\Omega_0)$ at which the field being dressed is located.
		For the dressing of an asymptotic massive particle falling into the black hole,
			we take the Wilson line to be along the particle's geodesic and take the limit $r_0\to 2M$.
		We will refer to this as the Wilson line puncture.
		}
	\label{fig:dressing}
\end{figure}

\begin{figure}[t]
	\centering
	\includegraphics[width=.9\textwidth]{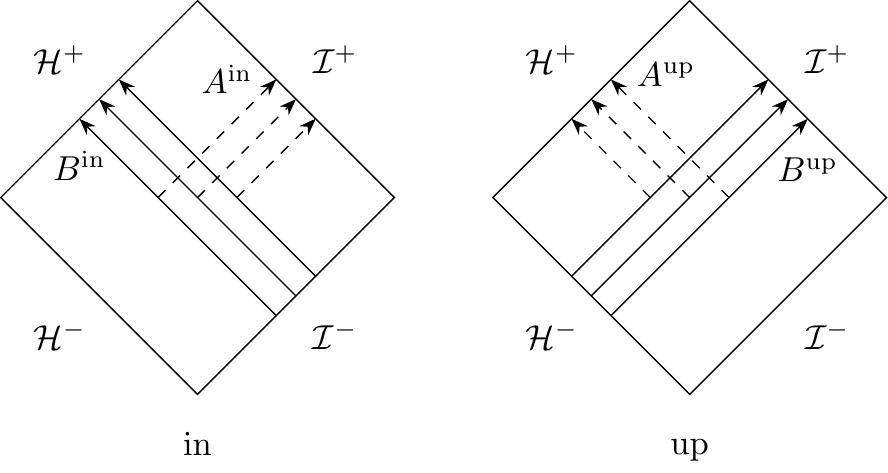}
	\caption{Representations of the ``in'' and ``up'' modes in the Penrose diagram of the exterior of a Schwarzschild black hole.
		The in-modes consist purely of traveling waves incoming from the past null infinity $\mI^-$ and therefore vanish on the past horizon $\mH^-$.
		The up-modes consist purely of traveling waves incoming from $\mH^-$ and therefore vanish on $\mI^-$.
		For each mode $\Lambda$, ${}_sR^\Lambda_{l\w}e^{-i\w t}$ is the incoming partial wave,
			$\left|A^\Lambda\right|^2$ is the reflection coefficient and $\left|B^\Lambda\right|^2$ is the transmission coefficient.}
	\label{fig:inup}
\end{figure}

In this section, we will apply the methods we reviewed in section \ref{sec:dressing} to construct
	the dressing of an asymptotic massive scalar field that falls into the Schwarzschild black hole.
Following Mandelstam's approach \cite{Mandelstam:1962us}, we assume that the scalar field is made coordinate-invariant
	by a gravitational Wilson line dressing, which now follows a time-like trajectory into the black hole.
After quantizing the graviton, we observe that the dressing comprises only zero-frequency graviton excitations.
It will then be shown in section \ref{sec:implant} that the dressing we construct carries a definite horizon supertranslation charge of Hawking, Perry and
	Strominger \cite{Hawking:2016msc,Hawking:2016sgy}.

To this end, let us first consider the dressing of a particle of mass $m$ at a spacetime point $z=(t_0,r_0,\theta_0,\phi_0)$.
Drawing analogy from section \ref{sec:dressing}, we write its dressing as the gravitational Wilson line,
\begin{align}
	\exp(W) \equiv \exp\left\{\frac{i}{2}m\kappa \int^z_\Gamma dx^\mu h_\mn(x)\frac{dx^\nu}{d\tau}\right\},
	\label{Wilson_line}
\end{align}
along a radial geodesic $\Gamma$ of a massive particle of mass $m$, extending from $z$ to the future horizon $\mH^+$,
	see figure \ref{fig:dressing}.
As usual, we employ the boundary condition \cite{Kulish:1970ut,Choi:2018oel} that the contribution to $W$ comes from
	only the upper bound, $z$, of the integral.
To obtain the dressing of an asymptotic particle on $\mH^+$, we evaluate the integral under the limit $r_0\to 2M$.
In this case, the entirety of the geodesic $\Gamma$ lies in the vicinity of $\mH^+$.
Since this Wilson line acts like a puncture on the future boundary $\mH^+_+$ of the horizon,
	we will refer to this as the Wilson line puncture, following \cite{Blommaert:2018oue, Choi:2018oel}.
	
We now employ the graviton quantization of Candelas et$.$ al$.$ \cite{Candelas:1981zv} (see appendix \ref{app:quant} for details),
	where the graviton field $h_\mn(x)$ has the mode expansion
\begin{align}
	h_\mn(x) = \sum_\Lambda \sum_{lmP}\int_0^\infty d\w \left[a^\Lambda_{lmP}(\w) h_\mn^\Lambda(l,m,\w,P;x) + \hc\right].
	\label{hmn_expansion}
\end{align}
The mode functions $h^\Lambda_\mn(l,m,\w,P;x)$ and their complex conjugates form a complete orthonormal set
	with $l\geq 2$, $|m|\leq l$, $P=\pm 1$, and
	$\Lambda\in\{\tin,\tup\}$.
Here $P=+1$ ($-1$) is referred to as the electric (magnetic) parity.
Modes with $\Lambda=\tin$ ($\tup$) are referred to as the in-modes (up-modes);
	it denotes the boundary conditions satisfied by the modes (see Fig. \ref{fig:inup}).
Throughout our paper, it will be tacitly assumed that the sum over $l$, $m$ and $P$ span $l\geq 2$, $|m|\leq l$ and $P=\pm 1$:
\begin{align}
	\sum_{lmP}\left(\cdots\right)\equiv \sum_{l\geq 2}\sum_{m=-l}^l\sum_{P=\pm 1}\left(\cdots\right),
\end{align}
unless explicitly stated otherwise.
The graviton is quantized by promoting $a^\Lambda_{lmP}(\w)$ and $a^{\Lambda\dagger}_{lmP}(\w)$
	to operators satisfying the canonical commutation relation \eqref{canonical}.

Coming back to the dressing \eqref{Wilson_line}, let us first consider the contribution to $W$ coming from the up-modes.\footnote{
	From figure \ref{fig:inup}, we see that the in-modes are incoming waves from $\mI^-$, which are known to have vanishing
		contribution to the horizon supertranslation charge (see \cite{Hawking:2016sgy} for a discussion).
	For this reason, we expect the dressing to receive vanishing contribution from the in-modes as well,
		and later in this section we will see that this is indeed the case.
}
First, we separate the graviton field \eqref{hmn_expansion} into two parts,
\begin{align}
	h_\mn(x) &= h^\tin_\mn(x) + h^\tup_\mn(x),\\
	h^\Lambda_\mn(x) &\equiv \sum_{lmP}\int_0^\infty d\w \left[a^\Lambda_{lmP}(\w) h_\mn^\Lambda(l,m,\w,P;x) + \hc\right],
		\quad \Lambda\in\{\tin,\tup\}.
\end{align}
Then, we may write \eqref{Wilson_line} as
\begin{align}
	W = \frac{i}{2}m\kappa \int^z_\Gamma dx^\mu h^\tup_\mn(x)\frac{dx^\nu}{d\tau}
		+ \text{(in-mode contribution)}.
	\label{firstsep}
\end{align}
Let $E$ be the total energy of the particle at infinity.
This fixes the geodesic $\Gamma$, along which we have
\begin{align}
	\frac{dt}{d\tau}=\frac{E}{mV},\qquad
	\frac{dr}{d\tau}=-\left(\frac{E^2}{m^2}-V\right)^{1/2}, \qquad
	\frac{d\theta}{d\tau}=\frac{d\phi}{d\tau}=0.
\end{align}
To simplify the calculations, we move to the ingoing Eddington-Finkelstein coordinates $(v,r,\theta,\phi)$,
	where we have $h^\tup_{vr}(x)=0$ and $h^\tup_{rr}(x)=0$.
Then \eqref{firstsep} simplifies to
\begin{align}
	W &=
		\frac{i}{2}m\kappa
			\int^{v_0} dv\,h^\tup_{vv}(x)\frac{dv}{d\tau}
			+ \text{(in-mode contribution)}
		\\ &=
		\frac{im^2\kappa}{4E}
			\int^{v_0} dv \,h^\tup_{vv}(x)
			+ \text{(in-mode contribution)}
		,
		\label{Wup}
\end{align}
where in the last equation we used
\begin{align}
	\frac{dv}{d\tau} = \frac{dt}{d\tau} + \frac{dr_*}{dr}\frac{dr}{d\tau}
		= \frac{E}{mV}\left[1 - \left(1-\frac{m^2V}{E^2}\right)^{1/2}\right]
		= \frac{m}{2E}+\mO(r-2M),
\end{align}
and discarded subleading terms in the expansion, since for an asymptotic particle $\Gamma$ lies entirely in the vicinity of $\mH^+$.
The component $h_{vv}^\tup(x)$ is a linear combination of the modes $h_{vv}^\tup(l,m,\w,P;x)$, whose explicit form may be read off from \eqref{up},
\begin{align}
	h_{vv}^\tup (l,m,\w,P;x) = N^\tup \left\{
			\Upsilon_{vv} \mmY(\theta,\phi) + P \Upsilon_{vv}^* \ppY(\theta,\phi)
		\right\}\ppRup(r)e^{-i \w t},
	\label{hvv}
\end{align}
where $N^\tup$ is a normalization constant, $\Upsilon_{vv}$ is a second-order differential operator defined as \eqref{Upsilon},
	${}_{\pm 2}Y_{lm}$ are $s=\pm 2$ spin-weighted spherical harmonics, and ${}_{+2}R_{l\w}^\tup$ is a radial function satisfying the
	boundary condition for up-modes; see appendix \ref{app:quant} for details.
Now, observe that \eqref{dh_deltaalpha} can be used to write the spin-weighted spherical harmonics in terms of ordinary spherical harmonics,
\begin{align}
	\Upsilon_\mn \mmY(\theta,\phi) &=
		-\frac{r^2V^2}{8}\dh\dh\mmY(\theta,\phi)
		=-\frac{r^2V^2}{8}\sqrt{\frac{(l+2)!}{(l-2)!}}\,Y_{lm}(\theta,\phi),
	\\
	\Upsilon_\mn^* \ppY(\theta,\phi) &=
		-\frac{r^2V^2}{8}\bar\dh\bar\dh\ppY(\theta,\phi)
		=-\frac{r^2V^2}{8}\sqrt{\frac{(l+2)!}{(l-2)!}}\,Y_{lm}(\theta,\phi).
\end{align}
Thus we may write \eqref{hvv} as
\begin{align}
	h_{vv}^\tup (l,m,\w,P;x) = -N^\tup (1+P)\frac{r^2V^2}{8}\sqrt{\frac{(l+2)!}{(l-2)!}}\,Y_{lm}(\theta,\phi)\ppRup(r)e^{-i \w t}.
\end{align}
One immediately sees that $h^\tup_{vv}(x)$ does not have $P=-1$ contribution,
\begin{align}
	h_{vv}^\tup(l,m,\w,P=-1;x)=0.
\end{align}
We will also see in section \ref{sec:st_charge} that supertranslation horizon charge only receives contribution from $P=1$ modes.
This is reminiscent of the gravitational memory at the asymptotic infinities (see for example \cite{Bieri:2013ada,Bieri:2018asm,Satishchandran:2019pyc}).
Since $\Gamma$ is near $\mH^+$, we can replace the radial function $\ppRup(r)$ by its asymptotic form \eqref{Rup} for $r\to 2M$,
\begin{align}
	\ppRup(r) \sim \frac{A^\tup_{l\w}}{(2M)^4V^2}e^{-i\w r_*} \qquad\text{near $\mH^+$},
\end{align}
which leads to
\begin{align}
	h_{vv}^\tup (l,m,\w,P;x) \sim
		-\frac{N^\tup A^\tup_{l\w}}{8(2M)^2}(1+P)\sqrt{\frac{(l+2)!}{(l-2)!}}\,Y_{lm}(\theta,\phi)e^{-i \w v}
		\qquad\text{near $\mH^+$.}
\end{align}
Expanding \eqref{Wup} into modes and substituting the above expression yields
\begin{align}
	W &=
		\frac{im^2\kappa}{4E}\sum_{lmP}\int_0^\infty d\w
			\int^{v_0} dv \left[
				a_{lmP}^\tup(\w)h^\tup_{vv}(l,m,\w,P;x)
				+ \hc
			\right]
		\nonumber\\&\qquad +\text{(in-mode contribution)}
		\\ &=
		-\frac{im^2\kappa}{4E}\sum_{lm}\sqrt{\frac{(l+2)!}{(l-2)!}}\int_0^\infty d\w
			\left[
				a_{lm,P=1}^\tup(\w)\frac{N^\tup A^\tup_{l\w}}{4(2M)^2}\,Y_{lm}(\theta,\phi)
					\frac{e^{-i \w v_0}}{(-i\w)}
				+ \hc
			\right]
		\nonumber\\&\qquad
			+ \text{(in-mode contribution)}
		,
\end{align}
where we have used a boundary condition analogous to that used in \cite{Kulish:1970ut, Choi:2018oel} to evaluate
\begin{align}
	\int^{v_0} dv\,e^{-i\w v} = \frac{e^{-i\w v_0}}{(-i\w)}.
\end{align}
Recall that the line integral was along a time-like geodesic $\Gamma$ of a particle with total energy $E$ at infinity,
	which implies that as $r_0\to 2M$, the advanced time $v_0$ diverges to infinity.
Thus in this limit, the presence of $e^{\pm i\w v_0}$ in the integrand removes all contributions except those from $\w=0$
	by virtue of the Riemann-Lebesgue lemma.
Following the previous approaches \cite{Kulish:1970ut, Ware:2013zja, Choi:2018oel}, we explicitly implement this
	by replacing $e^{\pm i\w v_0}$ with an infrared function $\phi(\w)$, which we define to have support only in a small neighborhood of $\w=0$
	and satisfy $\phi(0)=1$.
This yields
\begin{align}
	W &=
		-\frac{im^2\kappa}{4E}\sum_{lm}\sqrt{\frac{(l+2)!}{(l-2)!}}\int_0^\infty d\w \phi(\w)
			\left[
				a_{lm,P=1}^\tup(\w)\frac{N^\tup A^\tup_{l\w}}{(-4i\w)(2M)^2}\,Y_{lm}(\theta,\phi)
				+ \hc
			\right]
		\nonumber\\&\qquad
			+ \text{(in-mode contribution)}
		.
		\label{Wup2}
\end{align}
Due to the function $\phi(\w)$, only the leading soft term in the integrand contributes to the integral.
From \eqref{Nup} and \eqref{Aup}, we have the soft expansion
\begin{align}
	N^\tup A^\tup_{l\w} = (2M)^2\frac{(-4i\w)}{\sqrt{\pi\w}}\left[\frac{(l-2)!}{(l+2)!}+\mO(\w)\right].
\end{align}
We can substitute this to \eqref{Wup2} to obtain
\begin{align}
	W &=
		-\frac{im^2\kappa}{4E}\sum_{lm}\sqrt{\frac{(l-2)!}{(l+2)!}}\int_0^\infty \frac{d\w}{\sqrt{\pi\w}} \phi(\w)
			\left[
				a_{lm,P=1}^\tup(\w)\,Y_{lm}(\theta,\phi)
				+ \hc
			\right]
		\nonumber\\&\qquad
			+ \text{(in-mode contribution)}
		.
		\label{Wup3}
\end{align}

Now, let us turn our attention to last term in \eqref{Wup3}, the contribution from the in-modes.
One could imagine carrying out a similar set of steps, after which one would obtain an expression analogous to the first term
	in \eqref{Wup2}, where the integrand is proportional to $\phi(\w)N^\tin B^\tin_{l\w}$.
From \eqref{Nin} and \eqref{Bin}, we have the expansion
\begin{align}
	N^\tin B^\tin_{l\w} &=
		-\frac{(-4iM\w)^{l+1}}{2(2M)^3 \sqrt{\pi\w}}\left[\frac{l!(l-2)!(l+2)!}{2(2l+1)!(2l)!}+\mO(\w)\right],
\end{align}
which, when compared to \eqref{Wup2}, contains far more factors of $\w$ since $l\geq 2$.
This leads to the in-mode contribution being sub-leading soft in comparison to that of up-modes,
	and therefore negligible in comparison due to the presence of $\phi(\w)$.
There is a subtlety here that is worth mentioning: unlike the up-modes, the in-modes are in the ingoing radiation gauge $h_{(a)(1)}=0$, $g^\mn h_\mn=0$,
	which is not compatible with the Bondi gauge.
However, that the in-modes are sub-leading soft to the up-modes on $\mH^+$ is still true in Bondi gauge.
To see this, we note that the radial function $\mmRin(r)$ derives its origin from
	the contribution of each in-mode to the Weyl scalar $\Psi_4$ \cite{Jensen:1995qv, Candelas:1981zv},
\begin{align}
	\delta\Psi_4\left[h^\tin(l,m,\w,P;x)\right] &=
		-\frac{N^\tin}{8r^4}\left[\frac{(l+2)!}{(l-2)!} + 12iM\w P\right]\mmRin(r)\mmY(\theta,\phi)e^{-i\w t}.
	\label{Psi4}
\end{align}
Since $\delta\Psi_4$ is a gauge-invariant quantity, the Bondi gauge expression of the in-mode contribution
	to $W$ will also include the radial dependence $\mmRin(r)$, which includes the factor $\w^{l}$ near $\mH^+$,
	rendering the in-mode contribution sub-dominant in comparison to that of the up-modes.
Therefore, one obtains the final expression
\begin{align}
	\exp(W) &=
		\exp\left[
		-\frac{im^2\kappa}{4E}\sum_{lm}\sqrt{\frac{(l-2)!}{(l+2)!}}\int_0^\infty \frac{d\w}{\sqrt{\pi\w}} \phi(\w)
			\left[
				a_{lm,P=1}^\tup(\w)\,Y_{lm}(\theta,\phi)
				+ \hc
			\right]
		\right]
	\label{expW}
\end{align}
of the gravitational Wilson line, dressing an asymptotic particle of mass $m$ and total energy $E$ falling into the black hole.

We have seen that the in-modes are sub-leading soft to up-modes on $\mH^+$ and vanish by choice of boundary conditions on $\mH^-$.
Due to this observation, we can restrict our attention to the up-modes
	when dealing with supertranslation (and zero-modes) on both horizons.


\section{Supertranslation charge and horizon fields} \label{sec:st_charge}

Classical analysis of the black hole horizon \cite{Hawking:2016sgy} suggests that there exist horizon degrees of freedom of the form
	$h_{AB}$ that live on $\mH^\pm$.
Our goal in this section is to derive an expression of the horizon supertranslation charge and horizon fields in terms of
	the graviton Fock space operators.
From the previous work with regards to Maxwell fields \cite{Choi:2018oel}, it is reasonable to expect that such fields can be obtained as an appropriate limit
	of the bulk fields \eqref{hmn_expansion}.
However, one fails to do so directly on $\mH^+$, since $\ppRup$ blows up as one approaches $\mH^+$.
This is perhaps due to the fact that the bulk fields are Klein-Gordon normalized on $\mH^-\cup \mI^-$.
The limit, on the other hand, is well defined on $\mH^-$, which leads us to take an alternate approach:
	we derive the horizon fields on the past horizon $\mH^-$ first, and use time-inversion symmetry
	of Schwarzschild spacetime to obtain the corresponding fields on the future horizon $\mH^+$.

\subsection{Past horizon}

Let $\mH^- \cup \Sigma^-$ be a Cauchy surface in the past (for instance, in the absence of massive particles $\Sigma^- = \mI^-$).
The linearized supertranslation charge $Q_f^-$ on this surface can be decomposed as
\begin{align}
	Q_f^- = Q_f^{\mH^-} + Q_f^{\Sigma^-}.
\end{align}
We will loosely refer to $Q_f^{\mH^-}$ as the supertranslation charge on $\mH^-$.

To obtain an expression for $Q_f^{\mH^-}$, we move to the outgoing Eddington-Finkelstein coordinates $(u,r,\Omega)$,
	where $u=t-r_*$ is the retarded time.
The Schwarzschild metric in these coordinates reads
\begin{align}
	ds^2 = -Vdu^2-2dudr+r^2 \gamma_{AB}dx^Adx^B, \qquad V\equiv 1-\frac{2M}{r},
\end{align}
with the 2-sphere metric $\gamma_{AB}$.
The Bondi gauge conditions read \cite{Hawking:2016sgy}
\begin{align}
	h_{rr}=h_{rA}=\gamma^{AB}h_{AB}=0.
\end{align}
We want to find infinitesimal diffeomorphisms $\delta x^\mu=\xi^\mu$ that preserve the Bondi gauge conditions
	as well as the standard falloffs at large $r$ \cite{Hawking:2016sgy}.
Gauge conditions put the following constraints on $\xi^\mu$,
\begin{align}
	\frac{1}{2}\mL_\xi g_{rr} =\p_r\xi^u=0, \label{first}\\
	\mL_\xi g_{Ar}=\p_r\xi^A-\frac{1}{r^2}D^A f=0, \label{second}\\
	\frac{1}{2}\gamma^{AB}\mL_\xi g_{AB}=D_A\xi^A+\frac{2}{r}\xi^r=0. \label{third}
\end{align}
We restrict our attention to supertranslations by choosing $\xi^u=f$ such that $\p_u f=0$.
Then \eqref{first} leads to $f=f(\Omega)$, and \eqref{second} with falloff condition on $\xi^A$ implies $\xi^A = -\frac{1}{r}D^Af$.
Substituting this into \eqref{third}, one obtains $\xi^r = \frac{1}{2}D^2 f$.
Therefore we obtain
\begin{align}
	\xi^\alpha\p_\alpha=f\p_u + \frac{1}{2}D^2f\p_r - \frac{1}{r}D^A f\p_A.
\end{align}
In order to exclude ordinary spacetime translations which are not of our interest,
	we restrict the angular function $f(\Omega)$ to contain only partial waves with $l\geq 2$.

The supertranslation charge associated with the diffeomorphism $\xi$ on $\mH^-$ reads (see appendix \ref{app:Qminus} for a derivation)
\begin{align}
	Q^{\mH^-}_f =
		\frac{1}{\kappa M}
		\int_{\mH^-} d\Omega\,du\,f(\Omega)
			D^AD^B\p_u h^-_{AB}(u,\Omega),
	\label{Qminus}
\end{align}
where $h^-_{AB}(u,\Omega)$ are the horizon fields related to the supertranslation fields on $\mH^+$ obtained in \cite{Hawking:2016sgy};
	the ``$-$'' superscript emphasizes that these fields are defined on $\mH^-$.
The $u$-integral and $u$-derivative introduce a delta function $\delta(u)$ into the mode expansion of $h^-_{AB}$, which implies
	that it is only the zero-energy modes that are relevant for horizon supertranslation.

We can obtain the horizon fields $h_{AB}^-$ by taking the quantized graviton field $h_\mn(x)$ and taking the limit to $\mH^-$.
The boundary conditions are such that the in-modes vanish on $\mH^-$, so it suffices to consider the up-modes.
Although the up-modes are in the outgoing radiation gauge, the angular components $h^\tup_{AB}$
	already satisfy the Bondi gauge condition $\gamma^{AB}h_{AB}=0$ and therefore is expected to retain their functional form on $\mH^-$
	under a gauge transformation to Bondi gauge.
Recalling from \eqref{Rup} that
\begin{align}
	\ppRup(r)e^{-i\w t} \sim e^{-i\w u} \qquad \text{near $\mH^-$},
\end{align}
and observing from \eqref{Upsilon} that,
\begin{align}
	\Upsilon_{AB} &=
		- r^4 e_{(3)A}e_{(3)B}
		\left(
			\Delta
			+ 5\mu
			- 2\gamma
		\right)
		\left(
			\Delta
			+\mu
			- 4\gamma
		\right)
	\\ &=
		- r^4 e_{(3)A}e_{(3)B}
		\left(
			\p_u
			-\frac{V}{2}\p_r
			- \frac{5V}{2r}
			- \frac{M}{r^2}
		\right)
		\left(
			\p_u
			-\frac{V}{2}\p_r
			- \frac{V}{2r}
			- \frac{2M}{r^2}
		\right),
\end{align}
one obtains from \eqref{up} the asymptotic form
\begin{align}
	h^\text{up}_{AB}(l,m,\w,P;x) \sim
		-\frac{1}{2}(2M)^2 N^\text{up}
		H_{AB}(P;\Omega)
		e^{-i\w u} + \cdots
		\qquad\text{near $\mH^-$,}
	\label{hABup}
\end{align}
where $``\cdots"$ contains terms with additional factors of $\w$,
	which will be omitted since we're ultimately interested in leading soft modes.
Here we have defined
\begin{align}
	H_{AB}(P;\Omega) &\equiv
		\left.\left[
			e_{(3)A}e_{(3)B}\ {}_{-2}Y_{lm}(\Omega)
			+ Pe^*_{(3)A}e^*_{(3)B}\ {}_{+2}Y_{lm}(\Omega)
		\right]\right|_{r=2M}.
	\label{H_AB}
\end{align}
A tedious but straightforward computation shows that (see \ref{app:magnetic} for a derivation)
\begin{align}
	D^AD^BH_{AB}(P=-1;\Omega) = 0,
\end{align}
which implies
\begin{align}
	D^AD^B h^\tup_{AB}(l,m,\w,P=-1;x)=0,
\end{align}
that is, the magnetic parity modes $P=-1$ do not contribute to the supertranslation charge \eqref{Qminus}.
Again, this is similar to the fact that gravitational memory at the asymptotic infinities receive contribution only from the electric parity modes;
	see \cite{Bieri:2013ada,Bieri:2018asm,Satishchandran:2019pyc}.
For $P=1$, it can be shown that
\begin{align}
	H_{AB}(P=1;\Omega) = (2M)^2\sqrt{\frac{(l-2)!}{(l+2)!}}\left(2D_AD_B-\gamma_{AB}D^2\right)Y_{lm}(\Omega).
	\label{H_AB_P1}
\end{align}
Substituting the expressions to \eqref{hABup}, keeping only the relevant leading soft contribution
	and plugging the modes into the expansion \eqref{hmn_expansion} yields
\begin{align}
	h^-_{AB}(u,\Omega) &=
		-\frac{(2M)^4}{2}
			\left(2D_AD_B-\gamma_{AB}D^2\right)
			\nonumber\\&\qquad\times
			\sum_{lm} \sqrt{\frac{(l-2)!}{(l+2)!}}
			\int_0^\infty d\w 
			\left[
				N^\tup a^\tup_{lm,P=1}(\w)Y_{lm}(\Omega)e^{-i\w u}
				+ \hc
			\right].
		\label{hABu}
\end{align}
Since $D^2 Y_{lm}(\Omega) = -l(l+1)Y_{lm}(\Omega)$, and
\begin{align}
	D^AD^B(2D_AD_B-\gamma_{AB}D^2)Y_{lm}(\Omega) &=
		D^2(D^2+2)Y_{lm}(\Omega)
		=
		\frac{(l+2)!}{(l-2)!}Y_{lm}(\Omega),
	\label{D2}
\end{align}
we immediately obtain
\begin{align}
	D^AD^B\p_uh^-_{AB}(u,\Omega) &= 
		-\frac{(2M)^4}{2}
			\sum_{lm} \sqrt{\frac{(l+2)!}{(l-2)!}}
			\nonumber\\&\qquad\times
			\int_0^\infty d\w(-i\w) 
			\left[
				N^\tup a^\tup_{lm,P=1}(\w)Y_{lm}(\Omega)e^{-i\w u}
				- \hc
			\right].
		\label{DADBp_uh_AB}
\end{align}
Let us define the operator
\begin{align}
	N^-(\Omega) \equiv \frac{1}{\kappa M}\int_{-\infty}^\infty du\,D^AD^B\p_u h_{AB}^-(u,\Omega).
\end{align}
Using \eqref{DADBp_uh_AB} and the integral representation
\begin{align}
	\int_{-\infty}^\infty du\,e^{\pm i\w u} = 2\pi\delta(\w)
\end{align}
of the delta function, we obtain the expression
\begin{align}
	N^-(\Omega) &=
		\frac{2i\pi(2M)^3}{\kappa}
			\sum_{lm} \sqrt{\frac{(l+2)!}{(l-2)!}}
			\int_0^\infty d\w\,\w\,\delta(\w) 
			\left[
				N^\tup a^\tup_{lm,P=1}(\w)Y_{lm}(\Omega)
				- \hc
			\right].
		\label{Nm}
\end{align}
The supertranslation charge \eqref{Qminus} can be written in terms of the operator $N^-(\Omega)$ as
\begin{align}
	Q^{\mH^-}_f = \int d\Omega\,f(\Omega)N^-(\Omega).
	\label{QminusN}
\end{align}
The presence of the delta function $\delta(\w)$ clearly shows that only the zero-energy gravitons contribute
	to the supertranslation charge.

Now, let us take the horizon field $h_{AB}^-(u,\Omega)$ and take the limit $u\to \infty$,
	which brings the field to the future infinity $\mH^-_+$ of the past horizon.
Due to the factors $e^{\pm i\w u}$ in the integrand, by the the Riemann-Lebesgue lemma only the zero-energy modes will have
	non-vanishing contribution to the integral.
We can implement this by introducing the infrared function $\phi(\w)$ in place of $e^{\pm i\w u}$,
	which vanishes outside of a small neighborhood of $\w=0$ and satisfies $\phi(0)=1$.
Using this trick, we may define
\begin{align}
	h_{AB}^-(\Omega) &\equiv \lim_{u\to \infty}h^-_{AB}(u,\Omega)
		\\ &= 
			-2M\left(2D_AD_B-\gamma_{AB}D^2\right)\mA^-(\Omega)
			\label{hAB_Am}
\end{align}
where we introduced the scalar field
\begin{align}
	\mA^-(\Omega) &=
		\frac{1}{2}(2M)^3 \sum_{lm} \sqrt{\frac{(l-2)!}{(l+2)!}}
			\int_0^\infty d\w\,\phi(\w)
			\left[
				N^\tup a^\tup_{lm,P=1}(\w)Y_{lm}(\Omega)
				+ \hc
			\right],
	\label{Am}
\end{align}
with the infrared function $\phi(\w)$.
The operators $N^-(\Omega)$ and $\kappa \mA^-(\Omega)$ satisfy the commutation relation\footnote{
	In deriving \eqref{NmCm}, we used a crossing relation similar to that used in the original quantization \cite{Candelas:1981zv}
		to avoid dealing with factors of $1/2$ coming from delta functions sitting on the boundary of integration domains;
		see appendix \ref{app:unfold}.
		This is similar to using conventions such as
		\begin{align}
			\int_0^\infty d\w f(\w)\delta(\w) = \frac{1}{2}f(0),
		\end{align}
		which were used in, for example, \cite{Gabai:2016kuf,Choi:2017bna,Choi:2018oel}.
	}
\begin{align}
	\left[N^-(\Omega'),\kappa \mA^-(\Omega)\right] &=
		i\pi(2M)^6
			\sum_{lm}\sum_{l'm'}
				\sqrt{\frac{(l'+2)!}{(l'-2)!}}
				\sqrt{\frac{(l-2)!}{(l+2)!}}
			\int_0^\infty d\w'd\w\,\w'\,\delta(\w')\phi(\w)
			\nonumber\\&\quad\times
			\left[
				N^{\tup'} a^\tup_{l'm',P=1}(\w')Y_{l'm'}(\Omega')
				- \hc
				,
				N^\tup a^\tup_{lm,P=1}(\w)Y_{lm}(\Omega)
				+ \hc
			\right]
		\\ &=
			i\sum_{l\geq 2}\sum_{m} Y_{lm}(\Omega')Y^*_{lm}(\Omega)
		\\ &= i\delta^{(2)}(\Omega-\Omega') + \text{($l=0,1$ terms)}.
		\label{NmCm}
\end{align}
Since $f(\Omega)$ does not contain partial waves with $l=0,1$, equations \eqref{QminusN}, \eqref{hAB_Am} and \eqref{NmCm}
	lead to the commutator
\begin{align}
	\left[Q_f^{\mH^-},\kappa h_{AB}^-(\Omega)\right] = -2iM(2D_AD_B-\gamma_{AB}D^2)f(\Omega),
\end{align}
which is the anticipated quantum action of supertranslation on $\mH^-$,
	reflecting the Lie derivative
\begin{align}
	\mL_\xi g_{AB}|_{\mH^-} = -2M(2D_AD_B-\gamma_{AB}D^2)f(\Omega)
\end{align}
of the metric.

\subsection{Future horizon}

Now that we have derived the supertranslation charge and the horizon fields on $\mH^-$, we use the time-reversal symmetry
	of the Schwarzschild spacetime to obtain analogous results on $\mH^+$.
The appropriate choice of coordinates is the ingoing Eddington-Finkelstein coordinates $(v,t,\Omega)$ with advanced time $v=t+r_*$,
	in which the Schwarzschild metric reads
\begin{align}
	ds^2 = -Vdv^2 + 2dvdr + r^2\gamma_{AB}dx^Adx^B.
\end{align}
Since there are the horizon degrees of freedom $h^-_{AB}(u,\Omega)$ on $\mH^-$, one should obtain their counterparts $h^+_{AB}(v,\Omega)$
	on $\mH^+$ by taking $t\to -t$, or equivalently $u\to -v$.
Applying this to \eqref{hABu}, we obtain the future horizon field to be
\begin{align}
	h^+_{AB}(v,\Omega) &= h^-_{AB}(-v,\Omega)
		\\ &=
		-\frac{(2M)^4}{2}
			\left(2D_AD_B-\gamma_{AB}D^2\right)
			\nonumber\\&\qquad\times
			\sum_{lm} \sqrt{\frac{(l-2)!}{(l+2)!}}
			\int_0^\infty d\w 
			\left[
				N^\tup a^\tup_{lm,P=1}(\w)Y_{lm}(\Omega)e^{i\w v}
				+ \hc
			\right].
		\label{hABv}
\end{align}
From \cite{Hawking:2016sgy}, we know that the vector field which generates supertranslation on $\mH^+$ is
\begin{align}
	\zeta^\alpha\p_\alpha=f\p_v - \frac{1}{2}D^2f\p_r + \frac{1}{r}D^A f\p_A,
\end{align}
and that the associated supertranslation charge on $\mH^+$ is
\begin{align}
	Q^{\mH^+}_f =
		\frac{1}{\kappa M}
		\int_{\mH^+} d\Omega\,dv\,f(\Omega)
			D^AD^B\p_v h^+_{AB}(v,\Omega).
	\label{Qplus}
\end{align}
Let us define the operator
\begin{align}
	N^+(\Omega) &\equiv \frac{1}{\kappa M}\int_{-\infty}^\infty dv\,D^AD^B\p_v h^+_{AB}(v,\Omega),
\end{align}
in terms of which the charge $Q^{\mH^+}_f$ has the simple form
\begin{align}
	Q^{\mH^+}_f = \int d\Omega\,f(\Omega)N^+(\Omega).
\end{align}
Similar to the derivation of \eqref{Nm}, we plug in the mode expansion \eqref{hABv} and use \eqref{D2} to obtain
\begin{align}
	N^+(\Omega) &=
		-\frac{2i\pi(2M)^3}{\kappa}
			\sum_{lm} \sqrt{\frac{(l+2)!}{(l-2)!}}
			\int_0^\infty d\w\,\w\,\delta(\w)
			\left[
				N^\tup a^\tup_{lm,P=1}(\w)Y_{lm}(\Omega)
				- \hc
			\right],
\end{align}
which, as expected, only involves zero-energy gravitons.

As in the case of $\mH^-$, we can obtain the zero-modes $h^+_{AB}(\Omega)$ by taking $h^+_{AB}(v,\Omega)$
	to the past boundary $\mH^+_-$ of the future horizon.
From \eqref{hABv}, we have
\begin{align}
	h^+_{AB}(\Omega) &\equiv \lim_{v\to -\infty} h^+_{AB}(v,\Omega)
		\\ &= 2M(2D_AD_B-\gamma_{AB}D^2)\mA^+(\Omega),
			\label{hABp_Ap}
\end{align}
where $\mA^+(\Omega)$ is the scalar field defined as
\begin{align}
	\mA^+(\Omega) &=
		-\frac{1}{2}(2M)^3
			\sum_{lm} \sqrt{\frac{(l-2)!}{(l+2)!}}
			\int_0^\infty d\w\,\phi(\w)
			\left[
				N^\tup a^\tup_{lm,P=1}(\w)Y_{lm}(\Omega)
				+ \hc
			\right]
	\label{Ap}
	\\ &=
		-\frac{1}{2}
			\sum_{lm} \sqrt{\frac{(l-2)!}{(l+2)!}}
			\int_0^\infty \frac{d\w}{\sqrt{\pi\w}}\phi(\w)
			\left[
				a^\tup_{lm,P=1}(\w)Y_{lm}(\Omega)
				+ \hc
			\right].
	\label{Ap2}
\end{align}
In the second equality we used \eqref{Nup}.
The field $h_{AB}^+(\Omega)$ are, up to a factor of $\kappa$, the supertranslation zero modes $\delta_fg_{AB}$ obtained in \cite{Hawking:2016sgy}.
The operators $N^+(\Omega)$ and $\kappa \mA^+(\Omega)$ satisfy a commutation relation similar to \eqref{NmCm},
\begin{align}
	\left[N^+(\Omega'),\kappa \mA^+(\Omega)\right] &= i\delta^{(2)}(\Omega-\Omega')+\text{($l=0,1$ terms),}
\end{align}
from which we obtain
\begin{align}
	\left[Q^{\mH^+}_f,\kappa \mA^+(\Omega)\right] &= if(\Omega)
	\label{Q_Ap}
	,\\
	\left[Q^{\mH^+}_f,\kappa h^+_{AB}(\Omega)\right] &= 2iM(2D_AD_B-\gamma_{AB}D^2)f(\Omega),
	\label{QhAB}
\end{align}
which is the anticipated quantum action of supertranslation on the metric perturbation,
	correctly reflecting the classical result \cite{Hawking:2016sgy} on $\mH^+$,
\begin{align}
	\mL_{\zeta}g_{AB}|_{\mH^+}=2M(2D_AD_B-\gamma_{AB}D^2)f.
\end{align}
Equation \eqref{QhAB} shows that the supertranslation zero modes $\kappa h^+_{AB}$,
	written as a linear combination of zero-frequency electric-parity up-mode gravitons,
	are the symplectic partners of the linearized horizon charge $Q^{\mH^+}_f$ that enlarge the horizon phase space,
	as anticipated from \cite{Hawking:2016sgy}.

\subsection{Comments}

We notice that the structures of the horizon fields and zero-modes on $\mH^\pm$ are very similar
	to those of the past/future null infinities $\mI^\pm$ that are extensively studied in the literature.
The commutator \eqref{Q_Ap} and its counterpart on $\mH^-$ suggest that the two scalar fields $\mA^\pm(\Omega)$
	are the analogs of the Goldstone boson modes on $\mI^\pm$ for asymptotically flat spacetimes \cite{He:2014cra}.
Recall that we obtained the horizon fields on $\mH^+$ from those on $\mH^-$ via time-inversion symmetry of Schwarzschild spacetime.
This was used to derive \eqref{Am} and \eqref{Ap}, from which one obtains
\begin{align}
	\mA^-(\Omega) = -\mA^+(\Omega).
\end{align}
This is reminiscent of the antipodal matching conditions of $\mI^\pm$ for Christodoulou-Klainerman spaces
	\cite{Christodoulou:1993uv,Strominger:2013jfa,He:2014cra}.

Also, we note that the relations \eqref{hAB_Am} and \eqref{hABp_Ap} suggest $\mA^\pm$ to be related to the horizon analogs
	of the ``scalar memory'' $T$ introduced in \cite{Satishchandran:2019pyc} at $\mI^\pm$ for asymptotically flat spacetimes.
Recall that magnetic parity modes dropped out in the construction of the dressing and the charge, which is reminiscent of
	the situation of gravitational memory at infinities.

\section{Gravitational dressing implants supertranslation charge}\label{sec:implant}

In section \ref{sec:H} we obtained the dressing $\exp(W)$ for a particle of mass $m$ with energy $E$ that falls into the
	black hole.
In this section, we show that this dressing carries a definite horizon supertranslation charge.

Comparing the expression \eqref{expW} for $\exp(W)$ with the expression \eqref{Ap2} for $\mA^+$, one immediately recognizes that
	the exponent $W$ is proportional to the operator $\mA^+$,
\begin{align}
	W=W(m,E,\Omega) = \frac{im^2}{2E}\kappa\mA^+(\Omega).
\end{align}
This with \eqref{Q_Ap} implies the commutation relation
\begin{align}
	\left[Q_f^{\mH^+},e^{W(m,E,\Omega)}\right] = -\frac{m^2}{2E}f(\Omega) e^{W(m,E,\Omega)}.
\end{align}
Recall from \eqref{expW} that the dressing $\exp(W)$ is written purely in terms of zero-energy gravitons
	and therefore carries no energy.
Given a Schwarzschild black hole state $\ket{M_0,0}$ of mass $M_0$ with zero soft supertranslation charge, i.e.
\begin{align}
	Q_f^{\mH^+}\ket{M_0,0} = 0,
\end{align}
one can use the dressings to obtain other Schwarzschild black hole states carrying non-zero soft supertranslation charge,
\begin{align}
	\ket{M_0,(m,E,\Omega)} &\equiv e^{W(m,E,\Omega)}\ket{M_0,0},\\
	Q_f^{\mH^+}\ket{M_0,(m,E,\Omega)} &= -\frac{m^2}{2E}f(\Omega)\ket{M_0,(m,E,\Omega)}.
\end{align}
Therefore, our derivation of the gravitational dressing provides an example in the quantum theory of the classical construction
	of the supertranslation hair in \cite{Hawking:2016sgy}.

\section{Discussion}\label{sec:discussion}

In this paper, we have explicitly shown how to construct soft supertranslation hair on the horizon
	of Schwarzschild black holes within a quantum field theoretical framework.
The essential ingredient was the construction of dressed states by attaching Wilson lines to the infalling scalar particles.
This perspective works for dressed states implanting hair both at $\mI^\pm$ and at the horizon.
Our quantization procedure implies that a crucial component in our construction of soft charges and Faddeev-Kulish dressings
	is the existence of an infinite red-shift surface.
At this surface the Killing vector $\partial_u$ associated with the time-translation symmetry of the background spacetime becomes null.
A massive particle approaching this surface will only make contact asymptotically at $u=\infty$
	and at this point its dressing carrying the soft charge only contains soft gravitons ($\omega=0$).
This can be seen by expanding the dressing in terms of plane waves $e^{\pm i\omega u}$;
	as $u \rightarrow \infty$ only soft modes contribute by virtue of the Riemann-Lebesgue lemma.
Thus, we can conclusively confirm that there is structure at the horizon and not just any null surface.
However, there is evidence that this particular example of supertranslation hair does not appear to have relevance to the black hole information paradox.
Evidence that Hawking radiation is not modified by soft hair implanted by supertranslating shock waves comes from \cite{Compere:2019rof}, 
	at least using the mechanisms analyzed therein.
Their results complement those of \cite{Javadinazhed:2018mle} obtained from the perspective of dressing states with soft hair
	where the authors showed that the spectrum of Hawking radiation (without backreaction) emitted in the Schwarzschild background is unchanged 
after including the dressing of asymptotic states with soft form factors.
The relevance of soft hair to black hole entropy is still an open question.

During the course of our work we have noticed an interesting connection with recent work on generalized global symmetries
	\cite{Gaiotto:2014kfa} (see also \cite{Grozdanov:2016tdf}).
Of particular relevance to us are the one-form symmetries discussed there.
One form symmetries arise from a conserved two form current.
The dual of this current integrated over a co-dimension 2 surface gives the conserved flux or the charge.
The conserved charge counts the number of ``electric field lines'' or, more generally, strings that puncture the co-dimension 2 surface.
The objects that are ``charged'' under the one form symmetries are the Wilson lines which create and destroy strings.
Of particular significance to us is the case of source-free electrodynamics in which case the 2 form current is $F^{\mu\nu}$.
As discussed in the paper, for the case of asymptotic symmetries we are interested in the soft charges.
Thus from the above considerations, we expect the following commutation relations between
	the generator of large gauge transformation (LGT) symmetry and the electric Wilson line to be:
	(See also, eq. (3.3) of  \cite{Gaiotto:2014kfa})
\begin{align}
[ Q_s, W_C] = W_C\theta(\Omega \cap C).
\end{align}
Here, the Wilson line is along $C$ and the $\theta$ function evaluates to $\pm 1$ if the surface (at infinity or the horizon) and the line $C$ intersect.
It is important to note here that the soft charge and the Wilson lines are evaluated only involving the zero modes --
	i.e., the punctures on the surfaces are the relevant objects in this connection.
In fact, this is exactly the commutation relation we have obtained earlier in \cite{Choi:2018oel}:
	See eq. (2.47) of that paper for the case of LGT at asymptotic infinity,
	eq. (3.57) for the case of soft hair on the Rindler horizon and eq. (4.56) for the Schwarzschild horizon.

The case of soft supertranslation hair on the Schwarzschild horizon is very similar to electrodynamics.
There is a vast literature on the construction, in general relativity, of two form currents that are conserved in the source free case
	(for a review, see \cite{AIHPA_1966__4_1_1_0}).
Once again we have a picture of gravitational field lines puncturing co-dimension 2 surfaces.
For the case of BMS supertranslations at $\mI^\pm$ and the ones at the horizon,
	we are interested in the soft charges which generate these transformations.
The soft charges will then have the expected commutation relations with the corresponding Wilson lines obtained in eq. (2.51)
	for BMS supertranslations and in eq. (5.2) for the case of the Schwarzschild horizon.

The Wilson line perspective provides new insight into the structure of the asymptotic symmetries.
It would be interesting to apply this to Yang-Mills theories and
	to the case of other types of soft charges \cite{Haco:2018ske,Haco:2019ggi} on the black hole horizons.
We leave investigations in these directions to future work.

\acknowledgments

We are very grateful to David Garfinkle, Malcolm Perry, Gautam Satishchandran and Bob Wald for invaluable discussions.
SC would also like to thank the organizers and participants of the Quantum Gravity and Quantum Information workshop at CERN in March 2019
	for useful comments and conversations on the project.
SC is supported by the Samsung Scholarship and the Leinweber Graduate Fellowship.

\appendix

\section{Notation and conventions} \label{app:nnc}

The following symbols will appear frequently throughout the paper:
\begin{align}
	V &= 1-\frac{2M}{r}, \qquad
	\kappa^2 = 32 \pi G.
\end{align}
We employ the mostly positive metric convention, for which the Schwarzschild metric $g_\mn$ in the usual coordinates reads
\begin{align}
	ds^2 &= -V dt^2 + \frac{dr^2}{V} + r^2 \gamma_{AB} dx^A dx^B,
\end{align}
where $\gamma_{AB}=r^{-2} g_{AB}$ denotes the 2-sphere metric.
Here capital Latin letters span the angular coordinates $A,B,\ldots=2,3$, as opposed to lower case greek letters $\mu,\nu,\ldots = 0,1,2,3$.
In the spherical coordinates $(\theta,\phi)$, $\gamma_{AB} = \mathrm{diag}(1,\sin^2\theta)$.
We will often used $\Omega$ to denote the angular variables collectively;
	for example $f(\Omega)=f(\theta,\phi)$ and $d\Omega = \sin\theta d\theta d\phi$.
$\nabla_\mu$ will be used to denote covariant derivative with respect to the Schwarzschild metric $g_\mn$;
	$D_A$ will be used to denote covariant derivative with respect to the 2-sphere metric $\gamma_{AB}$.

\subsection{Newman-Penrose formalism} \label{app:nnc1}

Here we reproduce some relevant details of the Newman-Penrose (NP) formalism used in the quantization of \cite{Candelas:1981zv,Jensen:1995qv}.
We will use $(a),(b),\ldots$ to denote tetrad indices in contrast to the tensor indices $\mu,\nu,\ldots$.

The components of the Kinnersley tetrad ${e_{(a)}}^\mu$ are, in the Schwarzschild coordinates $(t,r,\theta,\phi)$,
\begin{gather}
	{e_{(1)}}^\mu = \left(\frac{1}{V}, 1, 0, 0\right), \quad
	\quad {e_{(2)}}^\mu = \left(\frac{1}{2}, -\frac{V}{2}, 0, 0\right), \\
	\quad {e_{(3)}}^\mu = {e_{(4)}}^{\mu *} = \frac{1}{\sqrt 2 r}\left(0, 0, 1, \frac{i}{\sin \theta}\right).
\end{gather}
These satisfy the orthonormality condition
\begin{align*}
	g_\mn {e_{(a)}}^\mu{e_{(b)}}^\nu = \eta_{(a)(b)},
	\label{tetrads_ortho}
\end{align*}
where $g_\mn$ is the Schwarzschild metric and $\eta_{(a)(b)}$ is a constant symmetric matrix,
\begin{align}
	\eta_{(a)(b)} = \begin{bmatrix}
		0&-1&0&0\\-1&0&0&0\\0&0&0&\ 1\ \\0&0&\ 1\ &0
	\end{bmatrix}.
\end{align}
Tetrad indices $(a),(b),\ldots$ are lowered/raised by $\eta_{(a)(b)}$ and its inverse $\eta^{(a)(b)}$.
In terms of the Kinnersley tetrad, only one of the five independent Weyl scalars is non-zero,
\begin{align}
	\Psi_0 &\equiv -C_{(1)(3)(1)(3)}=0, \\
	\Psi_1 &\equiv -C_{(1)(2)(1)(3)}=0, \\
	\Psi_2 &\equiv -C_{(1)(3)(4)(2)}=\frac{M}{r^3}, \\
	\Psi_3 &\equiv -C_{(1)(2)(4)(2)}=0, \\
	\Psi_4 &\equiv -C_{(2)(4)(2)(4)}=0.
\end{align}
Here we used the notation $C_{(a)(b)(c)(d)}=C_\mnrs {e_{(a)}}^\mu {e_{(b)}}^\nu {e_{(c)}}^\rho {e_{(d)}}^\sigma$,
	where $C_\mnrs$ is the Weyl tensor of the Schwarzschild spacetime.
Furthermore, among the spin coefficients
\begin{align}
	\gamma_{(a)(b)(c)} \equiv \nabla_\mu{e_{(a)}}^\nu e_{(b)\nu} {e_{(c)}}^\mu,
\end{align}
all but the following vanish:
\begin{align}
	\rho &\equiv \gamma_{(3)(1)(4)} = -\frac{1}{r},\\
	\mu &\equiv \gamma_{(2)(4)(3)} = -\frac{V}{2r},\\
	\gamma &\equiv \frac{1}{2}\left (\gamma_{(2)(1)(2)}+\gamma_{(3)(4)(2)}\right ) = \frac{M}{2r^2},\\
	\alpha &\equiv \frac{1}{2}\left (\gamma_{(2)(1)(4)}+\gamma_{(3)(4)(4)}\right ) = -\frac{\cot \theta}{2\sqrt 2 r},\\
	\beta &\equiv \frac{1}{2}\left (\gamma_{(2)(1)(3)}+\gamma_{(3)(4)(3)}\right ) = \frac{\cot\theta}{2\sqrt 2 r}.
\end{align}
We will also make use of the tetrad operators (cf. \cite{chandrasekhar1983mathematical}),
\begin{align}
	{e_{(a)}}^\mu \p_\mu = (D, \Delta, \delta, \delta^*),
	\label{tetrad_ops}
\end{align}
which, in the Schwarzschild coordinates $(t,r,\theta,\phi)$, can be written out explicitly as
\begin{align}
	D = \frac{1}{V}\p_t + \p_r, \qquad
	\Delta = \frac{1}{2}\p_t - \frac{V}{2}\p_r,\qquad
	\delta = \frac{1}{\sqrt 2 r}\left(\p_\theta + \frac{i}{\sin \theta}\p_\phi\right).
\end{align}

\section{Quantizing metric perturbations of Schwarzschild}\label{app:quant}

In this appendix, we review the quantization of the linear perturbations of the Schwarzschild metric
	studied in \cite{Candelas:1981zv, Jensen:1995qv},
	which takes advantage of the simplifications given by the NP formalism.
Among the many conventions of the NP formalism,
	we follow those given in Appendix A of \cite{Jensen:1995qv}.
The quantization is done in two particular gauges called the ingoing and outgoing radiation gauge
	-- since we will be working in the Bondi gauge, it is worth noting that the outgoing radiation gauge satisfies the Bondi gauge conditions
	in the advance time coordinates.

For a Schwarzschild black hole of mass $\mbh$, the spacetime is described by the metric $g_\mn(x)$ with the line element
\begin{align}
	ds^2 = -V dt^2 + \frac{dr^2}{V} + r^2 d\theta^2 + r^2\sin^2\theta d\phi^2, \qquad V\equiv 1-\frac{2M}{r},
\end{align}
in the usual coordinates $(t, r, \theta,\phi)$, where $2M=2G\mbh$ is the Schwarzschild radius.
The appropriate choice of tetrad that reflects the symmetries of the Schwarzschild spacetime is the Kinnersley tetrad (see appendix \ref{app:nnc1}),
	defined as
\begin{align}
	{e_{(1)}}^\mu &= \left(\frac{1}{V}, 1, 0, 0\right), \\
	\quad {e_{(2)}}^\mu &= \left(\frac{1}{2}, -\frac{V}{2}, 0, 0\right), \\
	\quad {e_{(3)}}^\mu = {e_{(4)}}^{\mu *} &= \frac{1}{\sqrt 2 r}\left(0, 0, 1, \frac{i}{\sin \theta}\right).
\end{align}
In terms of the Kinnersley tetrads, all spin coefficients vanish except
\begin{align}
	\rho = -\frac{1}{r}, \quad
	\mu = -\frac{V}{2r^2}, \quad
	\gamma = \frac{M}{2r^2}, \quad
	-\alpha = \beta = \frac{\cot \theta}{2\sqrt 2 r}.
	\label{spincoeff}
\end{align}
We consider the perturbed metric $g'_\mn$ around the Schwarzschild background $g_\mn$,
\begin{align}
	g'_\mn(x) = g_\mn(x) + \kappa h_\mn(x),
\end{align}
where $\kappa^2 = 32\pi G$.
The complete set of modes is
\begin{align}
	\left\{h^\Lambda_\mn(l,m,\w,P;x),h^{\Lambda *}_\mn(l,m,\w,P;x)\right\}_{\Lambda,l,m,\w,P},
\end{align}
where $l\geq 2$, $|m|\leq l$, and $\Lambda\in\{\tin,\tup\}$
	indicates the boundary condition satisfied by the mode (see Fig. \ref{fig:inup}).
Each mode has a definite parity, labeled by $P=\pm 1$.
In the literature $P=+1$ and $P=-1$ are referred to as the electric and magnetic parities, respectively (see for example \cite{Chrzanowski:1975wv}).
The $\Lambda=\tin$ modes (henceforth the in-modes) have the form
\begin{align}
	h_\mn^\tin (l,m,\w,P;x)=N^\tin \left\{
			\Theta_\mn \ppY(\theta,\phi) + P \Theta_\mn^* \mmY(\theta,\phi)
		\right\}\mmRin(r)e^{-i\w t}
	\label{in}
\end{align}
in the ingoing radiation gauge $h_\mn {e_{(1)}}^\nu = 0$, $g^\mn h_\mn = 0$.
The $\Lambda=\tup$ modes (henceforth the up-modes) have the form
\begin{align}
	h_\mn^\tup (l,m,\w,P;x) = N^\tup \left\{
			\Upsilon_\mn \mmY(\theta,\phi) + P \Upsilon_\mn^* \ppY(\theta,\phi)
		\right\}\ppRup(r)e^{-i \w t}
	\label{up}
\end{align}
in the outgoing radiation gauge $h_\mn {e_{(2)}}^\nu = 0$, $g^\mn h_\mn = 0$.
Here $N^\Lambda$ are the normalization constants that are independent of the spacetime point $x$,
	and $\Theta_\mn$, $\Upsilon_\mn$ are second-order differential operators defined as\footnote{
	The authors of \cite{Dias:2009ex} argue that there is a typo in the expression for $\Upsilon_\mn$
		in the literature.
	The corrections proposed therein do not affect our results.
	}
\begin{align}
	\Theta_\mn &= -e_{(1)\mu}e_{(1)\nu}(\delta^*-2\alpha)(\delta^*-4\alpha)-e_{(4)\mu}e_{(4)\nu}(D-\rho)(D+3\rho)
		\nonumber\\&\quad
		+\frac{1}{2}\left(e_{(1)\mu}e_{(4)\nu}+e_{(4)\mu}e_{(1)\nu}\right)
		\left[
			D(\delta^*-4\alpha)+(\delta^*-4\alpha)(D+3\rho)
		\right],
	\label{Theta}
	\\
	\Upsilon_\mn &= r^4 \Big\{
		-e_{(2)\mu}e_{(2)\nu}(\delta-2\alpha)(\delta-4\alpha)-e_{(3)\mu}e_{(3)\nu}(\Delta+5\mu-2\gamma)(\Delta+\mu-4\gamma)
		\nonumber\\&\quad
		+\frac{1}{2}\left(e_{(2)\mu}e_{(3)\nu}+e_{(3)\mu}e_{(2)\nu}\right)
		\left[
			(\delta-4\alpha)(\Delta+\mu-4\gamma)+(\Delta+4\mu-4\gamma)(\delta-4\alpha)
		\right]
		\Big\},
	\label{Upsilon}
\end{align}
where $D$, $\Delta$ and $\delta$ are differential operators defined by the relation \cite{chandrasekhar1983mathematical}
\begin{align}
	{e_{(a)}}^\mu \p_\mu \equiv (D,\Delta,\delta,\delta^*).
\end{align}
The angular functions ${}_sY_{lm}(\theta,\phi)$ are spin-weighted spherical harmonics,
	whose relevant properties are spelled out in appendix \ref{app:swsh}.
The radial functions $\mmRin(r)$ and $\ppRup(r)$ are solutions to the ordinary differential equation
\begin{align}
	\Bigg[
		\frac{1}{\left[r(r-2M)\right]^{s}}\frac{d}{dr}&\left(\left[r(r-2M)\right]^{s+1}\frac{d}{dr}\right)
		\nonumber\\&\qquad
		+\frac{\w^2r^4+2is\w r^2(r-3M)}{r(r-2M)}
		-(l-s)(l+s+1)
	\Bigg]{}_sR_{l\w}(r)=0,
\end{align}
with $s=-2$ and $s=+2$ respectively, subject to the boundary conditions
\begin{align}
	\mmRin(r) &\sim
		\begin{cases}
			B_{l\w}^\tin (4M^2V)^2 e^{-i\w r_*} &\text{as $r\to 2M$},
			\\
			r^{-1}e^{-i\w r_*}+A^\tin_{l\w}r^3e^{+i\w r_*} &\text{as $r\to \infty$},
		\end{cases}
	\label{Rin}
	\\
	\ppRup(r) &\sim
		\begin{cases}
			A^\tup_{l\w}(4M^2V)^{-2}e^{-i\w r_*}+e^{+i\w r_*} &\text{as $r\to 2M$},
			\\
			B^\tup_{l\w}r^{-5}e^{+i\w r_*} &\text{as $r\to \infty$}.
		\end{cases}
	\label{Rup}
\end{align}
Here $r_* = r + 2M\ln(r/2M-1)$ is the tortoise coordinate, and $A^\Lambda_{l\w}$, $B^\Lambda_{l\w}$ are the
	reflection and transmission amplitudes respectively; see Fig. \ref{fig:inup}.
In particular, we note that $B^\tin_{l\w}$ and $A^\tup_{l\w}$ have the small-$\w$ expansion
\begin{align}
	B^\tin_{l\w} &= \frac{1}{(2M)^5}\frac{l!(l-2)!(l+2)!}{2(2l+1)!(2l)!}(-4iM\w)^{l+3} + \mO(\w^{l+4}),
	\label{Bin}
	\\
	A^\tup_{l\w} &= 2(2M)^4 \frac{(l-2)!}{(l+2)!}(-4iM\w) + \mO(\w^2),
	\label{Aup}
\end{align}
which will prove to be useful.
The normalization constants $N^\Lambda$ are fixed by the orthonormality condition
\begin{align}
	\braket{h^\Lambda(l,m,\w,P;x),h^{\Lambda'}(l',m',\w',P';x)}=\delta_{\Lambda\Lambda'}\delta_{ll'}\delta_{mm'}\delta(\w-\w')\delta_{PP'}.
\end{align}
The Klein-Gordon inner product $\braket{\cdot,\cdot}$ between two symmetric tensor fields $\psi_\ab$ and $\phi_\ab$ is defined as
\begin{align}
	\braket{\psi,\phi}=\frac{i}{2}\int_S d\Sigma^\mu
		\left(
			\psi^{\ab*}\nabla_\mu \bar\phi_\ab-\phi^\ab\nabla_\mu\bar\psi^*_\ab
			+ 2\bar\phi_{\alpha\mu}\nabla_\beta\bar\psi^{\ab*}
			- 2\bar\psi_{\alpha\mu}\nabla_\beta\bar\phi^\ab
		\right),
\end{align}
where $\bar\psi_\ab$, $\bar\phi_\ab$ are the trace-free parts of $\psi_\ab$, $\phi_\ab$, respectively,
	and $S$ is some Cauchy surface; see \cite{Candelas:1981zv} or \cite{Chrzanowski:1974nr} for the construction of the inner product.
Taking $S$ to be $\mH^-\cup \mI^-$, the normalization constants become\footnote{
	Note that the normalization is different from \cite{Jensen:1995qv} since we expand $g'_\mn = g_\mn + \kappa h_\mn$
		and quantize $h_\mn$ in order to give the graviton field a mass dimension.
	}
\begin{align}
	\left|N^\tin\right|^{-2} &= 64\pi\w^5,
	\label{Nin}\\
	\left|N^\tup\right|^{-2} &= (2M)^6\pi\w(1+4M^2\w^2)(1+16M^2\w^2).
	\label{Nup}
\end{align}

Having obtained the complete set of orthonormal modes,
	we can write the linear perturbation of the Schwarzschild background as the expansion
\begin{align}
	h_\mn(x) = \sum_\Lambda \sum_{lmP}\int_0^\infty d\w \left[a^\Lambda_{lmP}(\w) h_\mn^\Lambda(l,m,\w,P;x) + \hc\right]
\end{align}
and quantize the field by promoting $a^\Lambda_{lmP}(\w)$ and $a^{\Lambda\dagger}_{lmP}(\w)$ to operators that satisfy
	the commutation relations
\begin{align}
	\left[a^\Lambda_{lmP}(\w),a^{\Lambda'\dagger}_{l'm'P'}(\w')\right] &= \delta_{\Lambda\Lambda'}\delta_{ll'}\delta_{mm'}\delta(\w-\w')\delta_{PP'},
	\\
	\left[a^\Lambda_{lmP}(\w),a^{\Lambda'}_{l'm'P'}(\w')\right]&=0
	=\left[a^{\Lambda\dagger}_{lmP}(\w),a^{\Lambda'\dagger}_{l'm'P'}(\w')\right].
	\label{canonical}
\end{align}

A peculiar feature of this method of quantization is that the two modes (``in'' and ``up'') are in different gauges.
This will not cause problems for us, because the in-modes are, by definition, the linearized gravity waves sent in from $\mI^-$,
	and it is known that these waves carry zero supertranslation charge;
	see \cite{Hawking:2016sgy}, or \cite{Avery:2016zce} for a recent account.
One can also observe this directly from the soft expansion \eqref{Bin} of the black hole absorption amplitude $B^\tin_{l\w}$, which is proportional to
	$\w^{l+3}$.
This point will become relevant in section \ref{sec:st_charge} when we compute the contribution of the up-modes to the supertranslation charge.

It is noteworthy that the up-modes, which are in the outgoing radiation gauge, also satisfy the Bondi
	gauge conditions in the ingoing Eddington-Finkelstein coordinates $(v,r,\theta,\phi)$, where $v=t+r_*$.
To see this, we note that in these coordinates,
\begin{align}
	e_{(2)v} = -\frac{V}{2},\qquad e_{(2)r}=0,
\end{align}
which, when substituted into \eqref{up}, implies that
\begin{align}
	h^\tup_{vr}(l,m,\w,P;x)&=0,\\
	h^\tup_{rr}(l,m,\w,P;x)&=0, \label{Bondi1}\\
	h^\tup_{rA}(l,m,\w,P;x)&=0. \label{Bondi2}
\end{align}
Furthermore, the orthonormality \eqref{tetrads_ortho} implies that
\begin{align}
	\gamma^{AB}\Upsilon_{AB}\propto \gamma^{AB}e_{(3)A}e_{(3)B}= r^2g^\mn e_{(3)\mu}e_{(3)\nu}=0,
\end{align}
from which we obtain
\begin{align}
	\gamma^{AB}h_{AB}^\tup(l,m,\w,P;x)=0. \label{Bondi3}
\end{align}
Equations \eqref{Bondi1}, \eqref{Bondi2} and \eqref{Bondi3} are the Bondi gauge conditions in these coordinates \cite{Hawking:2016sgy},
	so as far as we're in the ingoing Eddington-Finkelstein coordinates, we can safely pretend that the up-modes are quantized in the Bondi gauge.

\section{Spin-weighted spherical harmonics} \label{app:swsh}

In this section, we review the relevant definition and properties of the spin-weighted spherical harmonics.
For more details we refer the reader to \cite{Goldberg:1966uu,Campbell:1971rm}.

The spin-weighted spherical harmonics ${}_sY_{lm}(\theta, \phi)$ are defined for integers $|s|\leq l$ by the equations
\begin{align}
	{}_0Y_{lm}(\theta, \phi) &= Y_{lm}(\theta, \phi),\\
	{}_{s+1}Y_{lm}(\theta, \phi) &= \left[(l-s)(l+s+1)\right]^{-1/2} \dh \,_sY_{lm}(\theta,\phi),\\
	{}_{s-1}Y_{lm}(\theta, \phi) &= -\left[(l-s)(l+s+1)\right]^{-1/2} \bar\dh \,_sY_{lm}(\theta,\phi),
\end{align}
where $Y_{lm}(\theta, \phi)$ are the ordinary spherical harmonics,
\begin{align}
	Y_{lm}(\theta, \phi) = \sqrt{\frac{(2l+1)}{4\pi}\frac{(l-m)!}{(l+m)!}}\,P_{l}^m(\cos\theta)e^{im\phi},
\end{align}
and $\dh$ ,$\bar\dh$ are operators which act on a function $\eta$ of spin-weight $s$ as
\begin{align}
	\dh \eta &= -\left(\frac{\p}{\p \theta} + \frac{i}{\sin \theta}\frac{\p}{\p\phi}-s\cot\theta\right)\eta, \\
	\bar\dh\eta&= -\left(\frac{\p}{\p \theta} - \frac{i}{\sin \theta}\frac{\p}{\p\phi}+s\cot\theta\right)\eta.
\end{align}
The spin-weighted spherical harmonics ${}_sY_{lm}$ form a complete orthonormal set for any function of $\theta$ and $\phi$ with spin-weight $s$:
\begin{gather}
	\int_0^{2\pi}d\phi\int_0^\pi \sin\theta\, d\theta\, {}_sY_{lm}(\theta,\phi){}_s Y_{l'm'}^*(\theta,\phi) = \delta_{ll'}\delta_{mm'}, \\
	\sum_{l= |s|}^\infty \sum_{m=-l}^l {}_sY_{lm}(\theta,\phi){}_sY_{lm}^*(\theta',\phi') =
		\delta(\cos \theta - \cos \theta')\delta(\phi-\phi').
\end{gather}

\section{Useful formulas}\label{app:useful}


\subsection{Christoffel symbols for Schwarzschild spacetime}

In the advanced Eddington-Finkelstein coordinates $(v,r,\Omega)$, the non-vanishing Christoffel symbols are
\begin{align}
\begin{split}
	&\Gamma^v_{vv}=\frac{M}{r^2}, \quad \Gamma^r_{vv}=\frac{MV}{r^2}, \quad \Gamma^r_{rv}=-\frac{M}{r^2}, \\
	\Gamma^v_{AB}=-r\gamma_{AB}&, \quad \Gamma^r_{AB}=-rV\gamma_{AB}, \quad
	\Gamma^A_{rB}=\frac{1}{r}\delta^A_B, \quad \Gamma^A_{BC}={}^{(2)}\Gamma^A_{BC},
\end{split}
\end{align}
where ${}^{(2)}\Gamma^A_{BC}$ denotes Christoffel symbols for a 2-sphere with metric $\gamma_{AB}$.
In the spherical coordinates ($\theta,\phi$), the non-vanishing components are
\begin{align}
	{}^{(2)}\Gamma^\theta_{\phi\phi}=-\sin\theta\cos\theta,\quad
	{}^{(2)}\Gamma^\phi_{\phi\theta}=\cot\theta.
\end{align}
In the retarded Eddington-Finkelstein coordinates $(u,r,\Omega)$, the non-vanishing symbols are
\begin{align}
\begin{split}
	&\Gamma^u_{uu}=-\frac{M}{r^2}, \quad \Gamma^r_{uu}=\frac{MV}{r^2}, \quad \Gamma^r_{ru}=\frac{M}{r^2}, \\
	\Gamma^u_{AB}=r\gamma_{AB}&, \quad \Gamma^r_{AB}=-rV\gamma_{AB}, \quad
	\Gamma^A_{rB}=\frac{1}{r}\delta^A_B, \quad \Gamma^A_{BC}={}^{(2)}\Gamma^A_{BC}.
\end{split}
\end{align}

\subsection{Spin-2 spherical harmonics} \label{app:swsh1}

In the NP formulation of Schwarzschild spacetime reviewed in section \ref{app:quant},
	the operators $\dh$, $\bar\dh$ can be expressed in terms of the tetrad operator $\delta$ \eqref{tetrad_ops}
	and the spin coefficient $\alpha$ \eqref{spincoeff}:
\begin{align}
\begin{split}
	\delta+2s\alpha &=
		\frac{1}{\sqrt 2r} \left(
			\p_\theta
			+\frac{i}{\sin\theta}\p_\phi
			-s\cot\theta
		\right)
		= -\frac{\dh}{\sqrt 2r},
	\\
	\delta^*-2s\alpha &=
		\frac{1}{\sqrt 2r} \left(
			\p_\theta
			-\frac{i}{\sin\theta}\p_\phi
			+s\cot\theta
		\right)
		= -\frac{\bar \dh}{\sqrt 2r}.
	\label{dh_deltaalpha}
\end{split}
\end{align}
These combinations appear for example in the definitions of the differential operators $\Theta_\mn$ \eqref{Theta} and $\Upsilon_\mn$ \eqref{Upsilon}.

There is a relation between ${}_{\pm 2}Y_{lm}(\theta,\phi)$ and a 2-sphere tensor of the form
\begin{align}
	(2D_AD_B - \gamma_{AB}D^2)Y_{lm}(\theta,\phi),
\end{align}
which we show below.
Each spin-2 spherical harmonics can be written as two spin operators acting on the ordinary spherical harmonics,
\begin{align}
	{}_{-2}Y_{lm}
		&= \sqrt{\frac{(l-2)!}{(l+2)!}}\, \bar\dh \bar\dh Y_{lm}
	\\ &=
		\sqrt{\frac{(l-2)!}{(l+2)!}}
		\left(
			\p_\theta^2
			+ \frac{2i\cos\theta}{\sin^2\theta}\p_\phi
			- \frac{2i}{\sin\theta}\p_\theta\p_\phi
			- \frac{1}{\sin^2\theta}\p_\phi^2
			- \cot\theta\p_\theta
		\right)
		Y_{lm},
\\
	{}_{+2}Y_{lm}
		&= \sqrt{\frac{(l-2)!}{(l+2)!}}\,\dh \dh Y_{lm}
	\\ &=
		\sqrt{\frac{(l-2)!}{(l+2)!}}
		\left(
			\p_\theta^2
			- \frac{2i\cos\theta}{\sin^2\theta}\p_\phi
			+ \frac{2i}{\sin\theta}\p_\theta\p_\phi
			- \frac{1}{\sin^2\theta}\p_\phi^2
			- \cot\theta\p_\theta
		\right)
		Y_{lm}.
\end{align}
Consider the following linear combinations,
\begin{align}
\begin{split}
	{}_{-2}Y_{lm} + {}_{+2}Y_{lm} &=
		2\sqrt{\frac{(l-2)!}{(l+2)!}}
		\left(
			\p_\theta^2
			- \frac{1}{\sin^2\theta}\p_\phi^2
			- \cot\theta\p_\theta
		\right)
		Y_{lm},
	\\
	{}_{-2}Y_{lm} - {}_{+2}Y_{lm} &=
		2\sqrt{\frac{(l-2)!}{(l+2)!}}
		\left(
			\frac{2i\cos\theta}{\sin^2\theta}\p_\phi
			-\frac{2i}{\sin\theta}\p_\theta\p_\phi
		\right)
		Y_{lm}.
	\label{YpYYmY}
\end{split}
\end{align}
The components and the trace of the 2-sphere tensor $D_AD_BY_{lm}$ are
\begin{align}
	D_\theta D_\theta Y_{lm} &= \p_\theta^2 Y_{lm},
	\\
	D_\theta D_\phi Y_{lm} &= \left(\p_\theta\p_\phi - \cot\theta \p_\phi \right)Y_{lm}
		= D_\phi D_\theta Y_{lm},
	\\
	D_\phi D_\phi Y_{lm} &= \left(\p_\phi^2 + \sin\theta\cos\theta \p_\theta \right)Y_{lm},
	\\
	D^2 Y_{lm} &= \gamma^{AB}D_AD_B Y_{lm}
		= \left(\p_\theta^2 + \frac{1}{\sin^2\theta}\p_\phi^2 + \cot\theta \p_\theta\right)Y_{lm}.
\end{align}
We can use these to write
\begin{align}
	(2D_\theta D_\theta -\gamma_{\theta\theta}D^2) Y_{lm} &=
		\left(\p_\theta^2 - \frac{1}{\sin^2\theta}\p_\phi^2 - \cot\theta\p_\theta \right)Y_{lm},
	\\
	(2D_\theta D_\phi -\gamma_{\theta\phi}D^2) Y_{lm} &= 2D_\theta D_\phi Y_{lm} =
		i\sin\theta\left(
			\frac{2i\cos\theta}{\sin^2\theta}\p_\phi
			-\frac{2i}{\sin\theta}\p_\theta\p_\phi
		\right)Y_{lm},
	\\
	(2D_\phi D_\phi -\gamma_{\phi\phi}D^2) Y_{lm} &=
		-\sin^2\theta\left(
			\p_\theta^2
			- \frac{1}{\sin^2\theta}\p_\phi^2 
			- \cot\theta\p_\theta 
		\right)Y_{lm}.
\end{align}
Comparison with \eqref{YpYYmY} yields the relations
\begin{align}
	{}_{-2}Y_{lm} + {}_{+2}Y_{lm} &=
		2\sqrt{\frac{(l-2)!}{(l+2)!}}\,
		(2D_\theta D_\theta-\gamma_{\theta\theta}D^2)Y_{lm}
	\\&=
		-\frac{2}{\sin^2\theta}
		\sqrt{\frac{(l-2)!}{(l+2)!}}\,
		(2D_\phi D_\phi - \gamma_{\phi\phi}D^2)Y_{lm},
	\\
	{}_{-2}Y_{lm} - {}_{+2}Y_{lm} &=
		-\frac{2i}{\sin\theta}
		\sqrt{\frac{(l-2)!}{(l+2)!}}\,
		(2D_\theta D_\phi - \gamma_{\theta\phi}D^2)Y_{lm}.
\end{align}

\section{Supertranslation charge on $\mH^-$} \label{app:Qminus}

To derive the supertranslation charge on the past horizon $\mH^-$, we move to the outgoing Eddington-Finkelstein coordinates
	$(u,r,\Omega)$, where $u=t-r_*$.
The Schwarzschild metric in these coordinates reads
\begin{align}
	ds^2 = -Vdu^2-2dudr+r^2 \gamma_{AB}dx^Adx^B,
\end{align}
with the 2-sphere metric $\gamma_{AB}$.
The non-zero inverse metric components are given by
\begin{align}
	g^{ur}=-1, \quad g^{rr}=V, \quad g^{AB}=\frac{1}{r^2}\gamma^{AB}.
\end{align}
The Bondi gauge conditions read \cite{Hawking:2016sgy}
\begin{align}
	h_{rr}=h_{rA}=\gamma^{AB}h_{AB}=0.
\end{align}
We want to find infinitesimal diffeomorphisms $\delta x^\mu=\xi^\mu$ that respect the Bondi gauge conditions as well
	as the falloffs at large $r$.
Bondi gauge conditions put the following constraints on $\xi^\mu$,
\begin{align}
	\nabla_r\xi_r=\p_r\xi^u=0, \label{first2}\\
	\nabla_A\xi_r+\nabla_r\xi_A=\p_r\xi^A-\frac{1}{r^2}D^A f=0, \label{second2}\\
	\gamma^{AB}\nabla_A\xi_B=D_A\xi^A+\frac{2}{r}\xi^r=0. \label{third2}
\end{align}
We restrict our attention to supertranslations by choosing $\xi^u=f$ such that $\p_u f=0$.
Then \eqref{first2} leads to $f=f(\Omega)$, and \eqref{second2} with falloff condition on $\xi^A$ implies $\xi^A = -\frac{1}{r}D^Af$.
Substituting this into \eqref{third2}, one obtains $\xi^r = \frac{1}{2}D^2 f$.
Therefore we obtain
\begin{align}
	\xi^\alpha\p_\alpha=f\p_u + \frac{1}{2}D^2f\p_r - \frac{1}{r}D^A f\p_A.
\end{align}
The supertranslation charge associated with the boundary $\p\Sigma$ of a Cauchy surface $\Sigma$ and diffeomorphism $\xi$ reads
\begin{align}
	Q^{\Sigma}_\xi = -\frac{2}{\kappa^2}\int_{\p\Sigma}*F,
\end{align}
where $F$ is a two-form with components
\begin{align}
	F_\mn &=
		\frac{1}{2}(\nabla_\mu\xi_\nu-\nabla_\nu\xi_\mu)h
		+ (\nabla_\mu {h^\alpha}_\nu-\nabla_\nu {h^\alpha}_\mu)\xi_\alpha
		+ (\nabla_\alpha\xi_\mu{h^\alpha}_\nu-\nabla_\alpha\xi_\nu{h^\alpha}_\mu)
		\\ &\quad
		+ (\nabla_\alpha{h^\alpha}_\mu\xi_\nu-\nabla_\alpha{h^\alpha}_\nu\xi_\mu)
		+ (\xi_\mu\nabla_\nu h-\xi_\nu\nabla_\mu h).
\end{align}
When $\p\Sigma$ is the boundary of the past horizon $\mH^-$, we may write
\begin{align}
	Q^{\mH^-}_\xi=-\frac{2(2M)^2}{\kappa^2} \left.\int d\Omega\, F_{ru}\right|^{\mH^-_+}_{\mH^-_-},
\end{align}
where $d\Omega = \sin\theta d\theta d\phi$.
The relevant component reads
\begin{align}
	F_{ru} &=
		\frac{\kappa}{2}(\nabla_r\xi_u-\nabla_u\xi_r)h
		+ \kappa(\nabla_r h_{\alpha u}-\nabla_u h_{\alpha r})\xi^\alpha
		+ \kappa(\nabla_\alpha\xi_r{h^\alpha}_u-\nabla_\alpha\xi_u{h^\alpha}_r)
		\nonumber\\ &\quad
		+ \kappa(\nabla_\alpha{h^\alpha}_r\xi_u-\nabla_\alpha{h^\alpha}_u\xi_r)
		+ \kappa(\xi_r\nabla_u h-\xi_u\nabla_r h).
\end{align}
In the Bondi gauge $h = g^\mn h_\mn = -2h_{ru}$, so at $r=2M$ the terms are given by
\begin{align}
	\frac{1}{2}(\nabla_r\xi_u-\nabla_u\xi_r)h &=
		\frac{f}{2M} h_{ru},
	\\
	(\nabla_r h_{\alpha u}-\nabla_u h_{\alpha r})\xi^\alpha &=
		f(\p_r h_{u u}-\p_u h_{u r}-\frac{1}{2M} h_{ru})
		+ D^Af(\frac{1}{4M^2} h_{Au}-\frac{1}{2M}\p_r h_{A u})
		\nonumber\\&\quad
		+ \frac{1}{2}D^2f\p_r h_{r u}
	\\
	\nabla_\alpha\xi_r{h^\alpha}_u-\nabla_\alpha\xi_u{h^\alpha}_r &=
		-\frac{f}{2M} h_{u r},
	\\
	\nabla_\alpha{h^\alpha}_r\xi_u-\nabla_\alpha{h^\alpha}_u\xi_r &=
			f\left(
				-\p_u h_{r u}
				-\p_r h_{u u}
				+\frac{1}{2M}h_{ru}
				+\frac{1}{4M^2}D^A h_{A u}
				-\frac{1}{M} h_{uu}
			\right)
			\nonumber\\&\quad
			+\frac{1}{2}D^2f\p_r h_{u r}
			+\frac{1}{2M}D^2 f h_{ur}
	\\
	\xi_r\nabla_u h-\xi_u\nabla_r h &=
		-D^2 f\p_r h_{ru}+2f\p_u h_{ru}.
\end{align}
Substituting these into $F_{ru}$ yields, up to a total derivative on $S^2$,
\begin{align}
	F_{ru} &=
		-\frac{\kappa}{2M}D^Af\p_r h_{A u}
		-\frac{\kappa f}{M} h_{uu}
		+\frac{\kappa}{2M}D^2 f h_{ur}
,
\end{align}
which leads to the following expression for the charge on $\mH^-$,
\begin{align}
	Q^{\mH^-}_f &=
		-\frac{2(2M)^2}{\kappa} \int d\Omega
		\left.\left(
			-\frac{1}{2M}D^Af\p_r h_{A u}
			-\frac{f}{M} h_{uu}
			+\frac{1}{2M}D^2 f h_{ur}
		\right)\right|^{\mH^-_+}_{\mH^-_-}
	\\ &=
		\frac{4M}{\kappa} \int d\Omega
		\,du\,f(\Omega)\p_u \left(
			- D^A\p_r h_{A u}
			+ 2 h_{uu}
			- D^2 h_{ur}
		\right)
		\label{charge}
.
\end{align}
Given the linearly perturbed metric
\begin{align}
	g'_\mn = g_\mn + \kappa h_\mn,
\end{align}
the perturbed Ricci tensor is given by \cite{tHooft:1974toh}
\begin{align}
	R'_\mn = R_\mn
		- \frac{\kappa}{2} \left(
			\nabla_\nu\nabla_\mu h
			- \nabla_\rho\nabla_\mu {h_\nu}^\rho
			- \nabla_\rho\nabla_\nu {h_\mu}^\rho
			+ \nabla_\rho\nabla^\rho h_\mn
		\right)
		+ \mO(\kappa^2).
\end{align}
We will take the background $g_\mn$ to be the Schwarzschild metric,
	and use the obvious notation that quantities with primes are computed from the perturbed metric $g'_\mn$
	and those without are computed from $g_\mn$.
We keep everything only up to linear order in $h_\mn$, and indices are raised/lowered using $g_\mn$.
The non-trivial constraints on $\mH^-$ are,
\begin{align}
	G'_{uu}&= \frac{\kappa^2}{4} T'_{uu} = 0,
	\\
	G'_{uA}&= \frac{\kappa^2}{4} T'_{uA} = 0,
\end{align}
where $G'_\mn=R'_\mn-\frac{1}{2}g'_\mn R'$ is the perturbed Einstein tensor.
The perturbed energy-momentum tensor $T'_\mn$ vanishes since we keep terms only up to linear order in $h_\mn$.
At $r=2M$ we have
\begin{align}
	-\frac{1}{2}\nabla_u\nabla_u h &=
		\p_u^2 h_{ru}
		+ \frac{1}{4M} \p_u h_{ru}
, \\
	-\frac{1}{2}\nabla_\rho \nabla^\rho h_{uu} &=
		\p_r\p_u h_{uu}
		+ \frac{1}{2M} \p_u h_{uu}
		+ \frac{1}{4M} \p_r h_{uu}
		- \frac{1}{8M^2} D^2 h_{uu}
		- \frac{1}{2M} \p_u h_{ru}
		- \frac{1}{8M^2} h_{ru}
, \\
	\nabla_\rho\nabla_u {h_u}^\rho &=
		- \p_u^2 h_{ur}
		- \p_r\p_u h_{uu}
		+ \frac{1}{4M^2}\p_u D^A h_{uA}
		+ \frac{1}{4M} \p_u h_{ru}
		- \frac{1}{M}  \p_u h_{uu}
		- \frac{1}{4M} \p_r h_{uu}
		\nonumber\\&\quad
		+ \frac{1}{8M^2} h_{ru}
		+ \frac{1}{16M^3} D^A h_{uA}
.
\end{align}
The constraint $G'_{uu}=0$ thus reduces to
\begin{align}
	0 &= \frac{4M^2}{\kappa} G'_{uu} = 4M^2\left(
			- \frac{1}{2}\nabla_u\nabla_u h
			- \frac{1}{2}\nabla_\rho\nabla^\rho h_{uu}
			+ \nabla_\rho\nabla_u {h_u}^\rho
		\right)
		\\ &=
		\p_u (D^A h_{uA} - 2M h_{uu})
		+ \frac{1}{4M} D^A h_{uA}
		- \frac{1}{2} D^2 h_{uu}
	.
\end{align}
We can use the following expressions at $r=2M$,
\begin{align}
	\nabla_\rho\nabla_A {h_u}^\rho &=
		- \p_r D_A h_{uu}
		- \p_u D_A h_{ur}
		+ \frac{1}{2M} D_A h_{ru}
		- \frac{1}{M} D_A h_{uu}
		- \frac{3}{8M^2} h_{uA}
		\nonumber\\&\quad
		+ \frac{1}{4M^2} D^BD_A h_{uB}
, \\
	\nabla_\rho\nabla_u {h_A}^\rho &=
		- \p_u \p_r h_{Au}
		- \frac{1}{M} \p_u h_{uA}
		- \frac{1}{8M^2} h_{uA}
		+ \frac{1}{4M^2}\p_u D^B h_{AB}
, \\
	- \nabla_u\nabla_A h &=
		2 \p_u D_A h_{ur}
, \\
	- \nabla_\rho\nabla^\rho h_{Au} &=
		2 \p_u \p_r h_{Au}
		- \frac{1}{4M^2} D^2 h_{Au}
		+ \frac{1}{M} D_A h_{uu}
,
\end{align}
to write the constraint $G'_{Au}=R'_{Au}=0$ as
\begin{align}
	0 &= \frac{2}{\kappa} G'_{Au} =
		\nabla_\rho\nabla_A {h_u}^\rho
		+ \nabla_\rho\nabla_u {h_A}^\rho
		- \nabla_u\nabla_A h
		- \nabla_\rho\nabla^\rho h_{Au}.
	\\ &=
		\p_u \left(
			D_A h_{ur}
			+ \p_r h_{Au}
			- \frac{1}{M} h_{uA}
			+ \frac{1}{4M^2} D^B h_{AB}
		\right)
		- D_A \p_r h_{uu}
		+ \frac{1}{2M} D_A h_{ru}
		\nonumber\\&\quad
		+ \frac{1}{4M^2} D_AD^B h_{uB}
		- \frac{1}{4M^2} D^2 h_{Au}
		- \frac{1}{4M^2} h_{uA}
\end{align}
Taking the linear combination $0=\frac{4M}{\kappa}G'_{uu}+\frac{2}{\kappa}D^AG'_{Au}$, we obtain the equation
\begin{align}
	\p_u \left(
		- D^A \p_r h_{Au}
		+ 2 h_{uu}
		- D^2 h_{ur}
	\right) &=
		\frac{1}{4M^2} D^AD^B\p_u h_{AB}
		- \frac{1}{2M} D^2 h_{uu}
		\nonumber\\&\quad
		- D^2 \p_r h_{uu}
		+ \frac{1}{2M} D^2 h_{ru}
		- \frac{1}{4M^2} D^A h_{uA}
	.
\end{align}
Substituting this back into \eqref{charge} yields the following expression for the horizon charge,
\begin{align}
	Q^{\mH^-}_f =\frac{1}{\kappa M}
		\int d\Omega\,du\,f(\Omega)
		&\bigg(
			D^AD^B\p_u h_{AB}
			- D^A h_{uA}
			\nonumber\\&\quad
			- 2M D^2 h_{uu}
			- 4M^2 D^2 \p_r h_{uu}
			+ 2M D^2 h_{ru}
		\bigg)
.
\end{align}
By performing a gauge fixing analogous to \cite{Hawking:2016sgy}, the expression reduces to
\begin{align}
	Q^{\mH^-}_f =
		\frac{1}{\kappa M}
		\int d\Omega\,du\,f(\Omega)
			D^AD^B\p_u h_{AB}
	.
	\label{charge2}
\end{align}

\section{Magnetic parity contribution}\label{app:magnetic}

Here we show that the magnetic parity $P=-1$ does not contribute to the supertranslation charge \eqref{Qminus} on $\mH^-$.
By the choice of boundary conditions, the in-modes vanish on $\mH^-$, so it suffices to consider the up-modes.

Recall from \eqref{hABup} that $D^AD^B h^\tup_{AB}$ is proportional to the following quantity,
\begin{align}
	D^AD^B H_{AB} &=
		D_\theta D_\theta H_{\theta\theta}
		+ \frac{1}{\sin^2\theta} D_\theta D_\phi H_{\theta\phi}
		+ \frac{1}{\sin^2\theta} D_\phi D_\theta H_{\phi\theta}
		+ \frac{1}{\sin^4\theta} D_\phi D_\phi H_{\phi\phi},
	\label{DADBHAB_MP}
\end{align}
where the tensor $H_{AB}$ is defined in \eqref{H_AB}.
Let us restrict our attention to the magnetic parity $P=-1$,
	for which the components of $H_{AB}$ reads
\begin{align}
	H_{\theta\theta}(P=-1;\Omega) &=
		2M^2 \left({}_{-2}Y_{lm} - {}_{+2}Y_{lm}\right),
	\\
	H_{\theta\phi}(P=-1;\Omega) &=
		2M^2 i\sin\theta \left({}_{-2}Y_{lm} + {}_{+2}Y_{lm}\right)
	\\
	H_{\phi\phi}(P=-1;\Omega) &=
		-2M^2\sin^2\theta \left({}_{-2}Y_{lm} - {}_{+2}Y_{lm}\right),
\end{align}
where we have, from appendix \ref{app:swsh1},
\begin{align}
	{}_{-2}Y_{lm} + {}_{+2}Y_{lm} &=
		2\sqrt{\frac{(l-2)!}{(l+2)!}}
		\left(
			\p_\theta^2
			- \frac{1}{\sin^2\theta}\p_\phi^2
			- \cot\theta\p_\theta
		\right)
		Y_{lm},
	\\
	{}_{-2}Y_{lm} - {}_{+2}Y_{lm} &=
		2\sqrt{\frac{(l-2)!}{(l+2)!}}
		\left(
			\frac{2i\cos\theta}{\sin^2\theta}\p_\phi
			-\frac{2i}{\sin\theta}\p_\theta\p_\phi
		\right)
		Y_{lm}.
\end{align}
With some algebra, one can show that
\begin{align}
	D_\theta D_\theta H_{\theta\theta}(-1;\Omega) &=
		\frac{8iM^2}{\sin \theta}\sqrt{\frac{(l-2)!}{(l+2)!}}
		\bigg[
			\left(
				\cot^3 \theta
				+ \frac{5\cot\theta}{\sin^2\theta}
			\right) \p_\phi
		\nonumber\\&\quad
			-\p_\theta^3\p_\phi
			+ 3\left(
				1-\frac{2}{\sin^2\theta}
			\right)\p_\theta\p_\phi
			+ 3\cot\theta\p_\theta^2\p_\phi
		\bigg]
		Y_{lm}(\Omega)
		\label{DD1}
	, \\
	D_\theta D_\phi H_{\theta\phi}(-1;\Omega) &=
		\frac{-4iM^2}{\sin\theta}\sqrt{\frac{(l-2)!}{(l+2)!}}\bigg[
			\frac{11\cos\theta+\cos 3\theta}{\sin\theta}\p_\phi
			-3\cot\theta\p_\phi^3
			-\sin^2\theta\p_\theta^3\p_\phi
			\nonumber\\&\quad
			+\p_\theta\p_\phi^3
			-\frac{1}{2}(19+9\cos 2\theta)\p_\theta\p_\phi
			+3\sin 2\theta\p_\theta^2\p_\phi
		\bigg]Y_{lm}(\Omega)
		\label{DD2}
	,
\end{align}
and
\begin{align}
	D_\phi D_\theta H_{\theta\phi}(-1;\Omega) &=
		-4iM^2\sin\theta\sqrt{\frac{(l-2)!}{(l+2)!}}\bigg[
			13\p_\theta\p_\phi
			+\frac{10\cos\theta+2\cos 3\theta}{\sin^3\theta}\p_\phi
			-\frac{3\cot\theta}{\sin^2\theta}\p_\phi^3
		\nonumber\\&\quad
			-\p_\theta^3\p_\phi
			-\frac{14}{\sin^2\theta}\p_\theta\p_\phi
			+\frac{1}{\sin^2\theta}\p_\theta\p_\phi^3
			+6\cot\theta\p_\theta^2\p_\phi
		\bigg]Y_{lm}(\Omega)
		\label{DD3}
	, \\
	D_\phi D_\phi H_{\phi\phi}(-1;\Omega) &=
		8iM^2\sqrt{\frac{(l-2)!}{(l+2)!}}\bigg[
			\sin\theta\left(
				-8\cos^2\theta\p_\theta\p_\phi
				+\p_\theta\p_\phi^3
				+3\cos\theta\sin\theta\p_\theta^2\p_\phi
			\right)
		\nonumber\\&\quad
			+(\cos\theta+5\cos^3\theta)\p_\phi
			- 3\cos\theta\p_\phi^3
		\bigg]Y_{lm}(\Omega).
		\label{DD4}
\end{align}
Substituting \eqref{DD1}-\eqref{DD4} into \eqref{DADBHAB_MP} yields
\begin{align}
	D^AD^BH_{AB}(-1;\Omega)=0.
\end{align}
Therefore, with \eqref{hABup} we conclude
\begin{align}
	D^AD^B \p_u h^\tup_{AB}(l,m,\w,P=-1;x)=0\qquad \text{on $\mH^-$,}
\end{align}
which shows that the magnetic parity modes $P=-1$ do not contribute to the supertranslation charge.
This is similar to the situation of gravitational memory at the infinities of asymptotically flat spacetimes,
	see \cite{Bieri:2013ada,Bieri:2018asm,Satishchandran:2019pyc} for relevant discussions of the gravitational memory effects.

\section{Unfolding the energy integral}\label{app:unfold}

In order to avoid dealing with expressions of the form $\int_0^\infty d\w\,\delta(\w)$, let us expand
\begin{align}
	h_{AB}(x) = \sum_{\Lambda}\sum_{l,m,P} \int_{-\infty}^\infty d\w\, a^\Lambda_{lmP}(\w)
		h_{AB}^\Lambda(l,m,\w,P;x),
	\label{map1}
\end{align}
where $\Lambda\in\{\tin,\tup\}$, and a crossing relation analogous to that used in \cite{Candelas:1981zv} is used to write
\begin{align}
	h_{AB}^\Lambda(l,m,\w,P;x) = \left[h^{\Lambda}_{AB} (l,m,-\w,P;x)\right]^*\qquad \text{for $\w<0$.}
	\label{map2}
\end{align}
The commutator between operators $a^\Lambda_{lmP}(\w)$ becomes
\begin{align}
	\left [a^\Lambda_{lmP}(\w),a^{\Lambda'}_{l'm'P'}(\w')\right ] = \delta_{\Lambda\Lambda'}\delta_{PP'}\delta_{ll'}\delta_{mm'}\delta(\w+\w').
\end{align}
Recall that only the up-modes and the electric parity $P=1$ contribute to the supertranslation charge.
From \eqref{hABup} and \eqref{H_AB_P1}, the asymptotic expression
\begin{align}
	h_{AB}^\tup(l,m,\w,P=1;x) \sim -M\sqrt{\frac{(l-2)!}{\pi\w(l+2)!}}(2D_AD_B-\gamma_{AB}D^2)Y_{lm}(\Omega)e^{-i\w u}
\end{align}
holds near $\mH^-$ for $\w>0$.
We will omit the subleading soft modes for they are irrelevant for this discussion.
The crossing relation \eqref{map2} implies that the above expression can be extended as
\begin{align}
	h_{AB}^\tup(l,m,\w,P=1;x) \sim -M\sqrt{\frac{(l-2)!}{\pi|\w|(l+2)!}}(2D_AD_B-\gamma_{AB}D^2)\widetilde Y_{lm}(u,\Omega)e^{-i\w u}
	\label{hABup_extended}
\end{align}
near $\mH^-$, to apply both to positive and negative $\w$, where we defined
\begin{align}
	\widetilde Y_{lm}(\w,\Omega) \equiv
	\begin{cases}
		Y_{lm}(\Omega) & \text{for $\w>0$,}\\
		Y^*_{lm}(\Omega) & \text{for $\w<0$.}
	\end{cases}
\end{align}
Using \eqref{hABup_extended} and following an analogous set of steps to derive \eqref{Nm},
	we now obtain an alternate expression for $N^-$, which reads
\begin{align}
	N^-(\Omega) &=
		\frac{2i}{\kappa}\sum_{lm}\int_{-\infty}^{\infty}d\w\,a^\tup_{lm,P=1}(\w)\delta(\w)
			\sqrt{\frac{\pi|\w|(l+2)!}{(l-2)!}}\widetilde Y_{lm}(\w,\Omega).
\end{align}
Similarly, the zero-modes and the scalar fields are given by
\begin{align}
	h^-_{AB}(\Omega) &= -2M(2D_AD_B-\gamma_{AB}D^2)\mA^{-}(\Omega), \\
	\mA^{-}(\Omega) &= \frac{1}{2}\sum_{lm}\int^\infty_{-\infty}d\w\,\phi(\w) a^\tup_{lm,P=1}(\w)
		\sqrt{\frac{(l-2)!}{\pi|\w|(l+2)!}}\widetilde Y_{lm}(\w,\Omega).
\end{align}
Then by direct calculation,
\begin{align}
	[N^-(\Omega),\kappa\mA^{-}(\Omega')] &=
	i\sum_{lm}\sum_{l'm'}\int_{-\infty}^\infty d\w d\w'
	\phi(\w')\delta(\w)
	\sqrt{\frac{|\w| (l+2)!(l'-2)!}{|\w'| (l-2)!(l'+2)!}}
	\nonumber\\&\qquad\times
	\left[
		a^\tup_{lm,P=1}(\w)\widetilde Y_{lm}(\w,\Omega),
		a^\tup_{lm,P=1}(\w')\widetilde Y_{l'm'}(\w',\Omega')
	\right]
	\\ &=
	i\sum_{lm}\int_{-\infty}^\infty d\w d\w'
	\phi(\w')\delta(\w+\w')\delta(\w)
	\widetilde Y_{lm}(\w,\Omega)
	\widetilde Y_{lm}(\w,\Omega').
\end{align}
Since the delta function $\delta(\w+\w')$ is non-zero only when $\w \w'<0$, let us keep the Hermitian combination\footnote{
	This is similar to the construction of zero-modes as a Hermitian combination of $\w>0$ and $\w<0$ modes
	in \cite{He:2014cra}.} at $\w=0$,
\begin{align}
	\widetilde Y_{lm}(\w,\Omega)\widetilde Y_{lm}(\w,\Omega')\Big|_{\w=0}
	= \frac{1}{2}\left(Y_{lm}(\Omega)Y_{lm}^*(\Omega') + Y_{lm}^*(\Omega)Y_{lm}(\Omega')\right).
\end{align}
This leads to the commutator
\begin{align}
	[N(\Omega),\kappa \mA^{-}(\Omega')] &=
		i \sum_{l= 2}^\infty\sum_{m=-l}^l
		Y_{lm}(\Omega)
		Y^*_{lm}(\Omega')
	\\ &= i\delta^{(2)}(\Omega-\Omega') + \text{($l=0,1$ terms)},
\end{align}
where we used the completeness relation of spherical harmonics.

\bibliographystyle{jhep} 
\bibliography{ST_hair}

\providecommand{\href}[2]{#2}\begingroup\raggedright\begin{thebibliography}{10}

\bibitem{Bondi:1962px}
H.~Bondi, M.~G.~J. van~der Burg and A.~W.~K. Metzner, \emph{{Gravitational
  waves in general relativity. 7. Waves from axisymmetric isolated systems}},
  \href{http://dx.doi.org/10.1098/rspa.1962.0161}{\emph{Proc. Roy. Soc. Lond.}
  {\bf A269} (1962) 21--52}.

\bibitem{Sachs:1962wk}
R.~K. Sachs, \emph{{Gravitational waves in general relativity. 8. Waves in
  asymptotically flat space-times}},
  \href{http://dx.doi.org/10.1098/rspa.1962.0206}{\emph{Proc. Roy. Soc. Lond.}
  {\bf A270} (1962) 103--126}.

\bibitem{Strominger:2013jfa}
A.~Strominger, \emph{{On BMS invariance of gravitational scattering}},
  \href{http://dx.doi.org/10.1007/JHEP07(2014)152}{\emph{JHEP} {\bf 07} (2014)
  152}, [\href{https://arxiv.org/abs/1312.2229}{{\tt 1312.2229}}].

\bibitem{He:2014laa}
T.~He, V.~Lysov, P.~Mitra and A.~Strominger, \emph{{BMS supertranslations and
  Weinberg's soft graviton theorem}},
  \href{http://dx.doi.org/10.1007/JHEP05(2015)151}{\emph{JHEP} {\bf 05} (2015)
  151}, [\href{https://arxiv.org/abs/1401.7026}{{\tt 1401.7026}}].

\bibitem{Weinberg:1965nx}
S.~Weinberg, \emph{{Infrared photons and gravitons}},
  \href{http://dx.doi.org/10.1103/PhysRev.140.B516}{\emph{Phys. Rev.} {\bf 140}
  (1965) B516--B524}.

\bibitem{He:2014cra}
T.~He, P.~Mitra, A.~P. Porfyriadis and A.~Strominger, \emph{{New symmetries of
  massless QED}}, \href{http://dx.doi.org/10.1007/JHEP10(2014)112}{\emph{JHEP}
  {\bf 10} (2014) 112}, [\href{https://arxiv.org/abs/1407.3789}{{\tt
  1407.3789}}].

\bibitem{Kapec:2014opa}
D.~Kapec, V.~Lysov, S.~Pasterski and A.~Strominger, \emph{{Semiclassical
  Virasoro symmetry of the quantum gravity $ \mathcal{S}$-matrix}},
  \href{http://dx.doi.org/10.1007/JHEP08(2014)058}{\emph{JHEP} {\bf 08} (2014)
  058}, [\href{https://arxiv.org/abs/1406.3312}{{\tt 1406.3312}}].

\bibitem{Strominger:2013lka}
A.~Strominger, \emph{{Asymptotic symmetries of Yang-Mills theory}},
  \href{http://dx.doi.org/10.1007/JHEP07(2014)151}{\emph{JHEP} {\bf 07} (2014)
  151}, [\href{https://arxiv.org/abs/1308.0589}{{\tt 1308.0589}}].

\bibitem{Kapec:2015ena}
D.~Kapec, M.~Pate and A.~Strominger, \emph{{New Symmetries of QED}},
  \href{http://dx.doi.org/10.4310/ATMP.2017.v21.n7.a7}{\emph{Adv. Theor. Math.
  Phys.} {\bf 21} (2017) 1769--1785},
  [\href{https://arxiv.org/abs/1506.02906}{{\tt 1506.02906}}].

\bibitem{Cachazo:2014fwa}
F.~Cachazo and A.~Strominger, \emph{{Evidence for a New Soft Graviton
  Theorem}},  \href{https://arxiv.org/abs/1404.4091}{{\tt 1404.4091}}.

\bibitem{Lysov:2014csa}
V.~Lysov, S.~Pasterski and A.~Strominger, \emph{{Low’s subleading soft
  theorem as a symmetry of QED}},
  \href{http://dx.doi.org/10.1103/PhysRevLett.113.111601}{\emph{Phys. Rev.
  Lett.} {\bf 113} (2014) 111601}, [\href{https://arxiv.org/abs/1407.3814}{{\tt
  1407.3814}}].

\bibitem{Strominger:2014pwa}
A.~Strominger and A.~Zhiboedov, \emph{{Gravitational memory, BMS
  supertranslations and soft theorems}},
  \href{http://dx.doi.org/10.1007/JHEP01(2016)086}{\emph{JHEP} {\bf 01} (2016)
  086}, [\href{https://arxiv.org/abs/1411.5745}{{\tt 1411.5745}}].

\bibitem{Kapec:2014zla}
D.~Kapec, V.~Lysov and A.~Strominger, \emph{{Asymptotic Symmetries of Massless
  QED in Even Dimensions}},
  \href{http://dx.doi.org/10.4310/ATMP.2017.v21.n7.a6}{\emph{Adv. Theor. Math.
  Phys.} {\bf 21} (2017) 1747--1767},
  [\href{https://arxiv.org/abs/1412.2763}{{\tt 1412.2763}}].

\bibitem{Kapec:2015vwa}
D.~Kapec, V.~Lysov, S.~Pasterski and A.~Strominger, \emph{{Higher-dimensional
  supertranslations and Weinberg's soft graviton theorem}},
  \href{http://dx.doi.org/10.4310/AMSA.2017.v2.n1.a2}{\emph{Ann. Math. Sci.
  Appl.} {\bf 2} (2017) 69--94}, [\href{https://arxiv.org/abs/1502.07644}{{\tt
  1502.07644}}].

\bibitem{He:2015zea}
T.~He, P.~Mitra and A.~Strominger, \emph{{2D Kac-Moody symmetry of 4D
  Yang-Mills theory}},
  \href{http://dx.doi.org/10.1007/JHEP10(2016)137}{\emph{JHEP} {\bf 10} (2016)
  137}, [\href{https://arxiv.org/abs/1503.02663}{{\tt 1503.02663}}].

\bibitem{Strominger:2015bla}
A.~Strominger, \emph{{Magnetic Corrections to the Soft Photon Theorem}},
  \href{http://dx.doi.org/10.1103/PhysRevLett.116.031602}{\emph{Phys. Rev.
  Lett.} {\bf 116} (2016) 031602},
  [\href{https://arxiv.org/abs/1509.00543}{{\tt 1509.00543}}].

\bibitem{Dumitrescu:2015fej}
T.~T. Dumitrescu, T.~He, P.~Mitra and A.~Strominger,
  \emph{{Infinite-Dimensional Fermionic Symmetry in Supersymmetric Gauge
  Theories}},  \href{https://arxiv.org/abs/1511.07429}{{\tt 1511.07429}}.

\bibitem{Bianchi:2014gla}
M.~Bianchi, S.~He, Y.-t. Huang and C.~Wen, \emph{{More on Soft Theorems: Trees,
  Loops and Strings}},
  \href{http://dx.doi.org/10.1103/PhysRevD.92.065022}{\emph{Phys. Rev.} {\bf
  D92} (2015) 065022}, [\href{https://arxiv.org/abs/1406.5155}{{\tt
  1406.5155}}].

\bibitem{Campiglia:2014yka}
M.~Campiglia and A.~Laddha, \emph{{Asymptotic symmetries and subleading soft
  graviton theorem}},
  \href{http://dx.doi.org/10.1103/PhysRevD.90.124028}{\emph{Phys. Rev.} {\bf
  D90} (2014) 124028}, [\href{https://arxiv.org/abs/1408.2228}{{\tt
  1408.2228}}].

\bibitem{Campiglia:2015kxa}
M.~Campiglia and A.~Laddha, \emph{{Asymptotic symmetries of gravity and soft
  theorems for massive particles}},
  \href{http://dx.doi.org/10.1007/JHEP12(2015)094}{\emph{JHEP} {\bf 12} (2015)
  094}, [\href{https://arxiv.org/abs/1509.01406}{{\tt 1509.01406}}].

\bibitem{Campiglia:2015qka}
M.~Campiglia and A.~Laddha, \emph{{Asymptotic symmetries of QED and
  Weinberg’s soft photon theorem}},
  \href{http://dx.doi.org/10.1007/JHEP07(2015)115}{\emph{JHEP} {\bf 07} (2015)
  115}, [\href{https://arxiv.org/abs/1505.05346}{{\tt 1505.05346}}].

\bibitem{Campiglia:2015yka}
M.~Campiglia and A.~Laddha, \emph{{New symmetries for the gravitational
  S-matrix}}, \href{http://dx.doi.org/10.1007/JHEP04(2015)076}{\emph{JHEP} {\bf
  04} (2015) 076}, [\href{https://arxiv.org/abs/1502.02318}{{\tt 1502.02318}}].

\bibitem{Campiglia:2016jdj}
M.~Campiglia and A.~Laddha, \emph{{Sub-subleading soft gravitons: New
  symmetries of quantum gravity?}},
  \href{http://dx.doi.org/10.1016/j.physletb.2016.11.046}{\emph{Phys. Lett.}
  {\bf B764} (2017) 218--221}, [\href{https://arxiv.org/abs/1605.09094}{{\tt
  1605.09094}}].

\bibitem{Laddha:2017vfh}
A.~Laddha and P.~Mitra, \emph{{Asymptotic Symmetries and Subleading Soft Photon
  Theorem in Effective Field Theories}},
  \href{http://dx.doi.org/10.1007/JHEP05(2018)132}{\emph{JHEP} {\bf 05} (2018)
  132}, [\href{https://arxiv.org/abs/1709.03850}{{\tt 1709.03850}}].

\bibitem{Campiglia:2017mua}
M.~Campiglia and R.~Eyheralde, \emph{{Asymptotic $U(1)$ charges at spatial
  infinity}}, \href{http://dx.doi.org/10.1007/JHEP11(2017)168}{\emph{JHEP} {\bf
  11} (2017) 168}, [\href{https://arxiv.org/abs/1703.07884}{{\tt 1703.07884}}].

\bibitem{Campiglia:2017dpg}
M.~Campiglia, L.~Coito and S.~Mizera, \emph{{Can scalars have asymptotic
  symmetries?}},
  \href{http://dx.doi.org/10.1103/PhysRevD.97.046002}{\emph{Phys. Rev.} {\bf
  D97} (2018) 046002}, [\href{https://arxiv.org/abs/1703.07885}{{\tt
  1703.07885}}].

\bibitem{Campiglia:2017xkp}
M.~Campiglia and L.~Coito, \emph{{Asymptotic charges from soft scalars in even
  dimensions}}, \href{http://dx.doi.org/10.1103/PhysRevD.97.066009}{\emph{Phys.
  Rev.} {\bf D97} (2018) 066009}, [\href{https://arxiv.org/abs/1711.05773}{{\tt
  1711.05773}}].

\bibitem{Choi:2017bna}
S.~Choi, U.~Kol and R.~Akhoury, \emph{{Asymptotic Dynamics in Perturbative
  Quantum Gravity and BMS Supertranslations}},
  \href{http://dx.doi.org/10.1007/JHEP01(2018)142}{\emph{JHEP} {\bf 01} (2018)
  142}, [\href{https://arxiv.org/abs/1708.05717}{{\tt 1708.05717}}].

\bibitem{Choi:2017ylo}
S.~Choi and R.~Akhoury, \emph{{BMS supertranslation symmetry implies
  Faddeev-Kulish amplitudes}},
  \href{http://dx.doi.org/10.1007/JHEP02(2018)171}{\emph{JHEP} {\bf 02} (2018)
  171}, [\href{https://arxiv.org/abs/1712.04551}{{\tt 1712.04551}}].

\bibitem{Ashtekar:2018lor}
A.~Ashtekar, M.~Campiglia and A.~Laddha, \emph{{Null infinity, the BMS group
  and infrared issues}},
  \href{http://dx.doi.org/10.1007/s10714-018-2464-3}{\emph{Gen. Rel. Grav.}
  {\bf 50} (2018) 140--163}, [\href{https://arxiv.org/abs/1808.07093}{{\tt
  1808.07093}}].

\bibitem{Campiglia:2018see}
M.~Campiglia, L.~Freidel, F.~Hopfmueller and R.~M. Soni, \emph{{Scalar
  Asymptotic Charges and Dual Large Gauge Transformations}},
  \href{http://dx.doi.org/10.1007/JHEP04(2019)003}{\emph{JHEP} {\bf 04} (2019)
  003}, [\href{https://arxiv.org/abs/1810.04213}{{\tt 1810.04213}}].

\bibitem{Hirai:2018ijc}
H.~Hirai and S.~Sugishita, \emph{{Conservation Laws from Asymptotic Symmetry
  and Subleading Charges in QED}},
  \href{http://dx.doi.org/10.1007/JHEP07(2018)122}{\emph{JHEP} {\bf 07} (2018)
  122}, [\href{https://arxiv.org/abs/1805.05651}{{\tt 1805.05651}}].

\bibitem{Campiglia:2019wxe}
M.~Campiglia and A.~Laddha, \emph{{Loop Corrected Soft Photon Theorem as a Ward
  Identity}},  \href{https://arxiv.org/abs/1903.09133}{{\tt 1903.09133}}.

\bibitem{He:2019jjk}
T.~He and P.~Mitra, \emph{{Asymptotic Symmetries and Weinberg's Soft Photon
  Theorem in Mink$_{d+2}$}},  \href{https://arxiv.org/abs/1903.02608}{{\tt
  1903.02608}}.

\bibitem{He:2019pll}
T.~He and P.~Mitra, \emph{{Asymptotic Symmetries in $(d+2)$-Dimensional Gauge
  Theories}},  \href{https://arxiv.org/abs/1903.03607}{{\tt 1903.03607}}.

\bibitem{Choi:2019rlz}
S.~Choi and R.~Akhoury, \emph{{Subleading soft dressings of asymptotic states
  in QED and perturbative quantum gravity}},
  \href{http://dx.doi.org/10.1007/JHEP09(2019)031}{\emph{JHEP} {\bf 09} (2019)
  031}, [\href{https://arxiv.org/abs/1907.05438}{{\tt 1907.05438}}].

\bibitem{Kapec:2017tkm}
D.~Kapec, M.~Perry, A.-M. Raclariu and A.~Strominger, \emph{{Infrared
  Divergences in QED, Revisited}},
  \href{http://dx.doi.org/10.1103/PhysRevD.96.085002}{\emph{Phys. Rev.} {\bf
  D96} (2017) 085002}, [\href{https://arxiv.org/abs/1705.04311}{{\tt
  1705.04311}}].

\bibitem{Gabai:2016kuf}
B.~Gabai and A.~Sever, \emph{{Large gauge symmetries and asymptotic states in
  QED}}, \href{http://dx.doi.org/10.1007/JHEP12(2016)095}{\emph{JHEP} {\bf 12}
  (2016) 095}, [\href{https://arxiv.org/abs/1607.08599}{{\tt 1607.08599}}].

\bibitem{Kulish:1970ut}
P.~P. Kulish and L.~D. Faddeev, \emph{{Asymptotic conditions and infrared
  divergences in quantum electrodynamics}},
  \href{http://dx.doi.org/10.1007/BF01066485}{\emph{Theor. Math. Phys.} {\bf 4}
  (1970) 745}.

\bibitem{Ware:2013zja}
J.~Ware, R.~Saotome and R.~Akhoury, \emph{{Construction of an asymptotic S
  matrix for perturbative quantum gravity}},
  \href{http://dx.doi.org/10.1007/JHEP10(2013)159}{\emph{JHEP} {\bf 10} (2013)
  159}, [\href{https://arxiv.org/abs/1308.6285}{{\tt 1308.6285}}].

\bibitem{Chung:1965zza}
V.~Chung, \emph{{Infrared divergence in quantum electrodynamics}},
  \href{http://dx.doi.org/10.1103/PhysRev.140.B1110}{\emph{Phys. Rev.} {\bf
  140} (1965) B1110--B1122}.

\bibitem{Hawking:2016msc}
S.~W. Hawking, M.~J. Perry and A.~Strominger, \emph{{Soft hair on black
  holes}}, \href{http://dx.doi.org/10.1103/PhysRevLett.116.231301}{\emph{Phys.
  Rev. Lett.} {\bf 116} (2016) 231301},
  [\href{https://arxiv.org/abs/1601.00921}{{\tt 1601.00921}}].

\bibitem{Hawking:2016sgy}
S.~W. Hawking, M.~J. Perry and A.~Strominger, \emph{{Superrotation charge and
  supertranslation hair on black holes}},
  \href{http://dx.doi.org/10.1007/JHEP05(2017)161}{\emph{JHEP} {\bf 05} (2017)
  161}, [\href{https://arxiv.org/abs/1611.09175}{{\tt 1611.09175}}].

\bibitem{Hawking:1974sw}
S.~W. Hawking, \emph{{Particle creation by black holes}},
  \href{http://dx.doi.org/10.1007/BF02345020, 10.1007/BF01608497}{\emph{Commun.
  Math. Phys.} {\bf 43} (1975) 199--220}.

\bibitem{Hawking:1976ra}
S.~W. Hawking, \emph{{Breakdown of predictability in gravitational collapse}},
  \href{http://dx.doi.org/10.1103/PhysRevD.14.2460}{\emph{Phys. Rev.} {\bf D14}
  (1976) 2460--2473}.

\bibitem{Flanagan:2015pxa}
{\'E}.~{\'E}. Flanagan and D.~A. Nichols, \emph{{Conserved charges of the
  extended Bondi-Metzner-Sachs algebra}},
  \href{http://dx.doi.org/10.1103/PhysRevD.95.044002}{\emph{Phys. Rev.} {\bf
  D95} (2017) 044002}, [\href{https://arxiv.org/abs/1510.03386}{{\tt
  1510.03386}}].

\bibitem{Averin:2016ybl}
A.~Averin, G.~Dvali, C.~Gomez and D.~Lust, \emph{{Gravitational Black Hole Hair
  from Event Horizon Supertranslations}},
  \href{http://dx.doi.org/10.1007/JHEP06(2016)088}{\emph{JHEP} {\bf 06} (2016)
  088}, [\href{https://arxiv.org/abs/1601.03725}{{\tt 1601.03725}}].

\bibitem{Compere:2016jwb}
G.~Compère and J.~Long, \emph{{Vacua of the gravitational field}},
  \href{http://dx.doi.org/10.1007/JHEP07(2016)137}{\emph{JHEP} {\bf 07} (2016)
  137}, [\href{https://arxiv.org/abs/1601.04958}{{\tt 1601.04958}}].

\bibitem{Sheikh-Jabbari:2016lzm}
M.~M. Sheikh-Jabbari, \emph{{Residual diffeomorphisms and symplectic soft
  hairs: The need to refine strict statement of equivalence principle}},
  \href{http://dx.doi.org/10.1142/S0218271816440193}{\emph{Int. J. Mod. Phys.}
  {\bf D25} (2016) 1644019}, [\href{https://arxiv.org/abs/1603.07862}{{\tt
  1603.07862}}].

\bibitem{Baxter:2016nml}
J.~E. Baxter, \emph{{On the global existence of hairy black holes and solitons
  in anti-de Sitter Einstein–Yang–Mills theories with compact semisimple
  gauge groups}}, \href{http://dx.doi.org/10.1007/s10714-016-2126-2}{\emph{Gen.
  Rel. Grav.} {\bf 48} (2016) 133--43},
  [\href{https://arxiv.org/abs/1604.05012}{{\tt 1604.05012}}].

\bibitem{Compere:2016gwf}
G.~Compère, \emph{{Bulk supertranslation memories: a concept reshaping the
  vacua and black holes of general relativity}},
  \href{http://dx.doi.org/10.1142/S0218271816440065}{\emph{Int. J. Mod. Phys.}
  {\bf D25} (2016) 1644006}, [\href{https://arxiv.org/abs/1606.00377}{{\tt
  1606.00377}}].

\bibitem{Mao:2016pwq}
P.~Mao, X.~Wu and H.~Zhang, \emph{{Soft hairs on isolated horizon implanted by
  electromagnetic fields}},
  \href{http://dx.doi.org/10.1088/1361-6382/aa59da}{\emph{Class. Quant. Grav.}
  {\bf 34} (2017) 055003}, [\href{https://arxiv.org/abs/1606.03226}{{\tt
  1606.03226}}].

\bibitem{Averin:2016hhm}
A.~Averin, G.~Dvali, C.~Gomez and D.~Lust, \emph{{Goldstone origin of black
  hole hair from supertranslations and criticality}},
  \href{http://dx.doi.org/10.1142/S0217732316300457}{\emph{Mod. Phys. Lett.}
  {\bf A31} (2016) 1630045}, [\href{https://arxiv.org/abs/1606.06260}{{\tt
  1606.06260}}].

\bibitem{Cardoso:2016ryw}
V.~Cardoso and L.~Gualtieri, \emph{{Testing the black hole ‘no-hair’
  hypothesis}},
  \href{http://dx.doi.org/10.1088/0264-9381/33/17/174001}{\emph{Class. Quant.
  Grav.} {\bf 33} (2016) 174001}, [\href{https://arxiv.org/abs/1607.03133}{{\tt
  1607.03133}}].

\bibitem{Grumiller:2016kcp}
D.~Grumiller, A.~Perez, S.~Prohazka, D.~Tempo and R.~Troncoso, \emph{{Higher
  Spin Black Holes with Soft Hair}},
  \href{http://dx.doi.org/10.1007/JHEP10(2016)119}{\emph{JHEP} {\bf 10} (2016)
  119}, [\href{https://arxiv.org/abs/1607.05360}{{\tt 1607.05360}}].

\bibitem{Donnay:2016ejv}
L.~Donnay, G.~Giribet, H.~A. González and M.~Pino, \emph{{Extended Symmetries
  at the Black Hole Horizon}},
  \href{http://dx.doi.org/10.1007/JHEP09(2016)100}{\emph{JHEP} {\bf 09} (2016)
  100}, [\href{https://arxiv.org/abs/1607.05703}{{\tt 1607.05703}}].

\bibitem{Tamburini:2017dig}
F.~Tamburini, M.~De~Laurentis, I.~Licata and B.~Thidé, \emph{{Twisted soft
  photon hair implants on Black Holes}},
  \href{http://dx.doi.org/10.3390/e19090458}{\emph{Entropy} {\bf 19} (2017)
  458}, [\href{https://arxiv.org/abs/1702.04094}{{\tt 1702.04094}}].

\bibitem{Ammon:2017vwt}
M.~Ammon, D.~Grumiller, S.~Prohazka, M.~Riegler and R.~Wutte,
  \emph{{Higher-Spin Flat Space Cosmologies with Soft Hair}},
  \href{http://dx.doi.org/10.1007/JHEP05(2017)031}{\emph{JHEP} {\bf 05} (2017)
  031}, [\href{https://arxiv.org/abs/1703.02594}{{\tt 1703.02594}}].

\bibitem{Zhang:2017geq}
P.~M. Zhang, C.~Duval, G.~W. Gibbons and P.~A. Horvathy, \emph{{Soft gravitons
  and the memory effect for plane gravitational waves}},
  \href{http://dx.doi.org/10.1103/PhysRevD.96.064013}{\emph{Phys. Rev.} {\bf
  D96} (2017) 064013}, [\href{https://arxiv.org/abs/1705.01378}{{\tt
  1705.01378}}].

\bibitem{Mishra:2017zan}
R.~K. Mishra and R.~Sundrum, \emph{{Asymptotic Symmetries, Holography and
  Topological Hair}},
  \href{http://dx.doi.org/10.1007/JHEP01(2018)014}{\emph{JHEP} {\bf 01} (2018)
  014}, [\href{https://arxiv.org/abs/1706.09080}{{\tt 1706.09080}}].

\bibitem{Gomez:2017ioy}
C.~Gomez and S.~Zell, \emph{{Black Hole Evaporation, Quantum Hair and
  Supertranslations}},
  \href{http://dx.doi.org/10.1140/epjc/s10052-018-5799-8}{\emph{Eur. Phys. J.}
  {\bf C78} (2018) 320}, [\href{https://arxiv.org/abs/1707.08580}{{\tt
  1707.08580}}].

\bibitem{Grumiller:2017otl}
D.~Grumiller, P.~Hacker and W.~Merbis, \emph{{Soft hairy warped black hole
  entropy}}, \href{http://dx.doi.org/10.1007/JHEP02(2018)010}{\emph{JHEP} {\bf
  02} (2018) 010}, [\href{https://arxiv.org/abs/1711.07975}{{\tt 1711.07975}}].

\bibitem{Chu:2018tzu}
C.-S. Chu and Y.~Koyama, \emph{{Soft Hair of Dynamical Black Hole and Hawking
  Radiation}}, \href{http://dx.doi.org/10.1007/JHEP04(2018)056}{\emph{JHEP}
  {\bf 04} (2018) 056}, [\href{https://arxiv.org/abs/1801.03658}{{\tt
  1801.03658}}].

\bibitem{Kirklin:2018wvq}
J.~Kirklin, \emph{{Localisation of Soft Charges, and Thermodynamics of Softly
  Hairy Black Holes}},
  \href{http://dx.doi.org/10.1088/1361-6382/aad204}{\emph{Class. Quant. Grav.}
  {\bf 35} (2018) 175010}, [\href{https://arxiv.org/abs/1802.08145}{{\tt
  1802.08145}}].

\bibitem{Cvetkovic:2018dmq}
B.~Cvetković and D.~Simić, \emph{{Near horizon OTT black hole asymptotic
  symmetries and soft hair}},
  \href{http://dx.doi.org/10.1088/1674-1137/43/1/013109}{\emph{Chin. Phys.}
  {\bf C43} (2019) 013109}, [\href{https://arxiv.org/abs/1804.00484}{{\tt
  1804.00484}}].

\bibitem{Grumiller:2018scv}
D.~Grumiller and M.~M. Sheikh-Jabbari, \emph{{Membrane Paradigm from Near
  Horizon Soft Hair}},
  \href{http://dx.doi.org/10.1142/S0218271818470065}{\emph{Int. J. Mod. Phys.}
  {\bf D27} (2018) 1847006}, [\href{https://arxiv.org/abs/1805.11099}{{\tt
  1805.11099}}].

\bibitem{Chandrasekaran:2018aop}
V.~Chandrasekaran, {\'E}.~{\'E}. Flanagan and K.~Prabhu, \emph{{Symmetries and
  charges of general relativity at null boundaries}},
  \href{http://dx.doi.org/10.1007/JHEP11(2018)125}{\emph{JHEP} {\bf 11} (2018)
  125}, [\href{https://arxiv.org/abs/1807.11499}{{\tt 1807.11499}}].

\bibitem{Averin:2018owq}
A.~Averin, \emph{{Schwarzschild/CFT from soft black hole hair?}},
  \href{http://dx.doi.org/10.1007/JHEP01(2019)092}{\emph{JHEP} {\bf 01} (2019)
  092}, [\href{https://arxiv.org/abs/1808.09923}{{\tt 1808.09923}}].

\bibitem{Donnay:2018ckb}
L.~Donnay, G.~Giribet, H.~A. González and A.~Puhm, \emph{{Black hole memory
  effect}}, \href{http://dx.doi.org/10.1103/PhysRevD.98.124016}{\emph{Phys.
  Rev.} {\bf D98} (2018) 124016}, [\href{https://arxiv.org/abs/1809.07266}{{\tt
  1809.07266}}].

\bibitem{Compere:2016hzt}
G.~Compère and J.~Long, \emph{{Classical static final state of collapse with
  supertranslation memory}},
  \href{http://dx.doi.org/10.1088/0264-9381/33/19/195001}{\emph{Class. Quant.
  Grav.} {\bf 33} (2016) 195001}, [\href{https://arxiv.org/abs/1602.05197}{{\tt
  1602.05197}}].

\bibitem{Donnay:2015abr}
L.~Donnay, G.~Giribet, H.~A. Gonzalez and M.~Pino, \emph{{Supertranslations and
  Superrotations at the Black Hole Horizon}},
  \href{http://dx.doi.org/10.1103/PhysRevLett.116.091101}{\emph{Phys. Rev.
  Lett.} {\bf 116} (2016) 091101},
  [\href{https://arxiv.org/abs/1511.08687}{{\tt 1511.08687}}].

\bibitem{Mandelstam:1962mi}
S.~Mandelstam, \emph{{Quantum electrodynamics without potentials}},
  \href{http://dx.doi.org/10.1016/0003-4916(62)90232-4}{\emph{Annals Phys.}
  {\bf 19} (1962) 1--24}.

\bibitem{Mandelstam:1962us}
S.~Mandelstam, \emph{{Quantization of the gravitational field}},
  \href{http://dx.doi.org/10.1016/0003-4916(62)90233-6}{\emph{Annals Phys.}
  {\bf 19} (1962) 25--66}.

\bibitem{Jakob:1990zi}
R.~Jakob and N.~G. Stefanis, \emph{{Path dependent phase factors and the
  infrared problem in QED}},
  \href{http://dx.doi.org/10.1016/0003-4916(91)90277-F}{\emph{Annals Phys.}
  {\bf 210} (1991) 112--136}.

\bibitem{Choi:2018oel}
S.~Choi and R.~Akhoury, \emph{{Soft photon hair on Schwarzschild horizon from a
  Wilson line perspective}},
  \href{http://dx.doi.org/10.1007/JHEP12(2018)074}{\emph{JHEP} {\bf 12} (2018)
  074}, [\href{https://arxiv.org/abs/1809.03467}{{\tt 1809.03467}}].

\bibitem{Candelas:1981zv}
P.~Candelas, P.~Chrzanowski and K.~W. Howard, \emph{{Quantization of
  electromagnetic and gravitational perturbations of a Kerr black hole}},
  \href{http://dx.doi.org/10.1103/PhysRevD.24.297}{\emph{Phys. Rev.} {\bf D24}
  (1981) 297--304}.

\bibitem{Jensen:1995qv}
B.~P. Jensen, J.~G. McLaughlin and A.~C. Ottewill, \emph{{One loop quantum
  gravity in Schwarzschild space-time}},
  \href{http://dx.doi.org/10.1103/PhysRevD.51.5676}{\emph{Phys. Rev.} {\bf D51}
  (1995) 5676--5697}, [\href{https://arxiv.org/abs/gr-qc/9412075}{{\tt
  gr-qc/9412075}}].

\bibitem{Gaiotto:2014kfa}
D.~Gaiotto, A.~Kapustin, N.~Seiberg and B.~Willett, \emph{{Generalized Global
  Symmetries}}, \href{http://dx.doi.org/10.1007/JHEP02(2015)172}{\emph{JHEP}
  {\bf 02} (2015) 172}, [\href{https://arxiv.org/abs/1412.5148}{{\tt
  1412.5148}}].

\bibitem{Satishchandran:2019pyc}
G.~Satishchandran and R.~M. Wald, \emph{{Asymptotic behavior of massless fields
  and the memory effect}},
  \href{http://dx.doi.org/10.1103/PhysRevD.99.084007}{\emph{Phys. Rev.} {\bf
  D99} (2019) 084007}, [\href{https://arxiv.org/abs/1901.05942}{{\tt
  1901.05942}}].

\bibitem{Donoghue:2017pgk}
J.~F. Donoghue, M.~M. Ivanov and A.~Shkerin, \emph{{EPFL lectures on general
  relativity as a quantum field theory}},
  \href{https://arxiv.org/abs/1702.00319}{{\tt 1702.00319}}.

\bibitem{Blommaert:2018oue}
A.~Blommaert, T.~G. Mertens and H.~Verschelde, \emph{{Edge dynamics from the
  path integral — Maxwell and Yang-Mills}},
  \href{http://dx.doi.org/10.1007/JHEP11(2018)080}{\emph{JHEP} {\bf 11} (2018)
  080}, [\href{https://arxiv.org/abs/1804.07585}{{\tt 1804.07585}}].

\bibitem{Blommaert:2018rsf}
A.~Blommaert, T.~G. Mertens, H.~Verschelde and V.~I. Zakharov, \emph{{Edge
  State Quantization: Vector Fields in Rindler}},
  \href{http://dx.doi.org/10.1007/JHEP08(2018)196}{\emph{JHEP} {\bf 08} (2018)
  196}, [\href{https://arxiv.org/abs/1801.09910}{{\tt 1801.09910}}].

\bibitem{Bieri:2013ada}
L.~Bieri and D.~Garfinkle, \emph{{Perturbative and gauge invariant treatment of
  gravitational wave memory}},
  \href{http://dx.doi.org/10.1103/PhysRevD.89.084039}{\emph{Phys. Rev.} {\bf
  D89} (2014) 084039}, [\href{https://arxiv.org/abs/1312.6871}{{\tt
  1312.6871}}].

\bibitem{Bieri:2018asm}
L.~Bieri, \emph{{Answering the Parity Question for Gravitational Wave Memory}},
  \href{http://dx.doi.org/10.1103/PhysRevD.98.124038}{\emph{Phys. Rev.} {\bf
  D98} (2018) 124038}, [\href{https://arxiv.org/abs/1811.09907}{{\tt
  1811.09907}}].

\bibitem{Christodoulou:1993uv}
D.~Christodoulou and S.~Klainerman, \emph{The global nonlinear stability of the
  minkowski space}, {\emph{S{\'e}minaire {\'E}quations aux d{\'e}riv{\'e}es
  partielles (Polytechnique)} (1990) 1--29}.

\bibitem{Compere:2019rof}
G.~Compère, J.~Long and M.~Riegler, \emph{{Invariance of Unruh and Hawking
  radiation under matter-induced supertranslations}},
  \href{http://dx.doi.org/10.1007/JHEP05(2019)053}{\emph{JHEP} {\bf 05} (2019)
  053}, [\href{https://arxiv.org/abs/1903.01812}{{\tt 1903.01812}}].

\bibitem{Javadinazhed:2018mle}
R.~Javadinezhad, U.~Kol and M.~Porrati, \emph{{Comments on Lorentz
  transformations, dressed asymptotic states and Hawking radiation}},
  \href{https://arxiv.org/abs/1808.02987}{{\tt 1808.02987}}.

\bibitem{Grozdanov:2016tdf}
S.~Grozdanov, D.~M. Hofman and N.~Iqbal, \emph{{Generalized global symmetries
  and dissipative magnetohydrodynamics}},
  \href{http://dx.doi.org/10.1103/PhysRevD.95.096003}{\emph{Phys. Rev.} {\bf
  D95} (2017) 096003}, [\href{https://arxiv.org/abs/1610.07392}{{\tt
  1610.07392}}].

\bibitem{AIHPA_1966__4_1_1_0}
C.~Cattaneo, \emph{Conservation laws}, {\emph{Annales de l'I.H.P. Physique
  th\'eorique} {\bf 4} (1966) 1--20}.

\bibitem{Haco:2018ske}
S.~Haco, S.~W. Hawking, M.~J. Perry and A.~Strominger, \emph{{Black Hole
  Entropy and Soft Hair}},
  \href{http://dx.doi.org/10.1007/JHEP12(2018)098}{\emph{JHEP} {\bf 12} (2018)
  098}, [\href{https://arxiv.org/abs/1810.01847}{{\tt 1810.01847}}].

\bibitem{Haco:2019ggi}
S.~Haco, M.~J. Perry and A.~Strominger, \emph{{Kerr-Newman Black Hole Entropy
  and Soft Hair}},  \href{https://arxiv.org/abs/1902.02247}{{\tt 1902.02247}}.

\bibitem{chandrasekhar1983mathematical}
S.~Chandrasekhar, \emph{The Mathematical Theory of Black Holes}.
\newblock The international series of monographs on physics. Clarendon Press,
  1983.

\bibitem{Chrzanowski:1975wv}
P.~L. Chrzanowski, \emph{{Vector potential and metric perturbations of a
  rotating black hole}},
  \href{http://dx.doi.org/10.1103/PhysRevD.11.2042}{\emph{Phys. Rev.} {\bf D11}
  (1975) 2042--2062}.

\bibitem{Dias:2009ex}
O.~J.~C. Dias, H.~S. Reall and J.~E. Santos, \emph{{Kerr-CFT and gravitational
  perturbations}},
  \href{http://dx.doi.org/10.1088/1126-6708/2009/08/101}{\emph{JHEP} {\bf 08}
  (2009) 101}, [\href{https://arxiv.org/abs/0906.2380}{{\tt 0906.2380}}].

\bibitem{Chrzanowski:1974nr}
P.~L. Chrzanowski and C.~W. Misner, \emph{{Geodesic synchrotron radiation in
  the Kerr geometry by the method of asymptotically factorized Green's
  functions}}, \href{http://dx.doi.org/10.1103/PhysRevD.10.1701}{\emph{Phys.
  Rev.} {\bf D10} (1974) 1701--1721}.

\bibitem{Avery:2016zce}
S.~G. Avery and B.~U.~W. Schwab, \emph{{Soft black hole absorption rates as
  conservation laws}},
  \href{http://dx.doi.org/10.1007/JHEP04(2017)053}{\emph{JHEP} {\bf 04} (2017)
  053}, [\href{https://arxiv.org/abs/1609.04397}{{\tt 1609.04397}}].

\bibitem{Goldberg:1966uu}
J.~N. Goldberg, A.~J. MacFarlane, E.~T. Newman, F.~Rohrlich and E.~C.~G.
  Sudarshan, \emph{{Spin s spherical harmonics and ð}},
  \href{http://dx.doi.org/10.1063/1.1705135}{\emph{J. Math. Phys.} {\bf 8}
  (1967) 2155}.

\bibitem{Campbell:1971rm}
W.~B. Campbell, \emph{{Tensor and spinor spherical harmonics and the spin-s
  harmonics ${}_sY_{lm}(\theta, \phi)$}},
  \href{http://dx.doi.org/10.1063/1.1665802}{\emph{J. Math. Phys.} {\bf 12}
  (1971) 1763--1770}.

\bibitem{tHooft:1974toh}
G.~'t~Hooft and M.~J.~G. Veltman, \emph{{One loop divergencies in the theory of
  gravitation}}, {\emph{Ann. Inst. H. Poincare Phys. Theor.} {\bf A20} (1974)
  69--94}.

\end{thebibliography}\endgroup

\end{document}